\documentclass{aa}  
\usepackage[utf8]{inputenc}
\usepackage[english]{babel}
\usepackage{enumitem}
\usepackage{appendix}

\usepackage{graphicx}
\usepackage{txfonts}

\usepackage{natbib}
\setcitestyle{authoryear, open{((},close{))}}

\usepackage{hyperref}
\hypersetup{
    colorlinks=true,
    linkcolor=blue,
    citecolor=blue,
    }

\def\teff{$\rm T_{\rm eff}$}
\def\gr{$\log {\rm g}$}
\def\vt{${\rm v_{\rm t}}$}

\begin{document} 
\title{A chemical close-up of the main body of the Sagittarius dwarf galaxy}
\author{
A. Liberatori\inst{1} \and
D. A. Alvarez Garay\inst{2} \and
M. Palla\inst{3,4} \and 
A. Mucciarelli\inst{3,4} \and 
M. Bellazzini\inst{4} \and
D. Romano\inst{4}
}

\institute{Section of Astrophysics, Astronomy and Mechanics, Physics Department, National and Kapodistrian University of Athens, Panepistimiopolis, 15784 Zografos, Athens, Greece
\and
INAF – Osservatorio Astrofisico di Arcetri, Largo E. Fermi 5, 50125 Firenze, Italy
\and
Dipartimento di Fisica e Astronomia “Augusto Righi", Alma Mater Studiorum, Università di Bologna, Via Gobetti 93/2, I-40129 Bologna, Italy
\and
INAF - Osservatorio di Astrofisica e Scienza dello Spazio di Bologna, Via Gobetti 93/3, 40129 Bologna, Italy
}

\date{}

\abstract{
We present the chemical composition of a sample of 37 red giant branch (RGB) stars belonging to the main body of the remnant of the Sagittarius (Sgr) 
dwarf spheroidal galaxy. All stars were observed with the FLAMES-UVES high-resolution spectrograph.
Twenty-three new targets are selected along the blue side of the RGB of Sgr, but outside the galaxy stellar nucleus, 
in order to avoid contamination by the stars of the metal-poor globular cluster M54. 
Additionally, we re-analyzed archival spectra of fourteen targets located on the red RGB.
For this sample, we derive the abundances of 21 chemical species (from Oxygen to Europium) representing different nucleosynthetic sites.
The sample covers a large range of metallicity, from [Fe/H]$\sim -$2 to $\sim -$0.4 dex and we can identify the transition
between the enrichment phases dominated by CC-SNe and SNe Ia.
The observed [$\alpha$/Fe] trend suggests a knee occurring at [Fe/H]$\sim -$1.5/$-$1.3 dex, compatible with the rather low star formation efficiency of Sgr. At lower [Fe/H], Sgr stars exhibit a chemical composition compatible with Milky Way stars of similar [Fe/H]. 
The only relevant exceptions are [Mn/Fe], [Zn/Fe], and [Eu/Fe]. 
At [Fe/H] higher than $-$1.5/$-$1.3 dex, instead, the chemical pattern of Sgr significantly deviates from that of the Milky Way for almost 
all the elements analyzed in this study.
Some of the abundance patterns reveal a lower contribution by very massive stars exploding as hypernovae 
(e.g. [Mn/Fe], [Zn/Fe]), a higher contribution by sub-Chandrasekhar progenitors of SNe Ia (e.g. [Ni/Fe]) and a high production
efficiency of rapid neutron-capture elements ([Eu/Fe]).}


\maketitle
%

\section{Introduction}
\label{sec:intro}

The remnant of the Sagittarius dwarf spheroidal galaxy \citep[Sgr~dSph,][]{ibata1994dwarf} 
is the closest and most emblematic case of a dwarf satellite in an advanced stage of tidal disruption, due to 
its ongoing merging with a larger galaxy, the Milky Way (MW). 
According to the most recent models, Sgr experienced several perigalactic passages aroind the MW, with the first infall occurring around 5 Gyr ago \citep{ruiz2020recurrent}. 
Currently we observe a low-surface-brightness elongated spheroid remnant (the main body) and a tidal stream with leading and trailing branches for tens of kpc along their quasi-polar orbit \citep{majewski2003two,ibata20,ramos22}. 
By this interaction, Sgr is both actively influencing the star formation history of the MW and contributing to the formation 
of the MW halo by donating its stars \citep{laporte2019stellar,ruiz2020recurrent}.

Sgr~dSph has a stellar nucleus with a composite stellar population in terms of ages and metallicities, with a markedly bimodal metallicity distribution dominated by an old ($\simeq 10-13$~ Gyr) metal-poor component peaking around [Fe/H]$\simeq -1.6$ , corresponding to the globular cluster M~54, and an intermediate-age ($\simeq 4-6$~ Gyr) metal-rich component peaking around [Fe/H]$\simeq -0.5$, plus minor populations reaching super-solar metallicity at age $\simeq 1.0$~Gyr \citep{siegel07,bellazzini2008nucleus,alfaro2019}.
On the other hand, the core of the main body of Sgr outside the nucleus, exhibits a mono-modal metallicity 
distribution, with a metal-rich peak at [Fe/H]$\simeq -0.5$, as in the nucleus, and a weak, extended metal-poor tail 
reaching [Fe/H]$\le -2.0$ \citep[][and references therein]{minelli2023metallicity}. It is generally accepted that the progenitor of Sgr~dSph displayed a significant radial gradient in metallicity and, likely, in age, that today is traced both in the main body and in the tidal stream \citep{bellazzini1999,alard2001, bellazzini06,monaco07stream,carlin12,de2014alpha,deboer15x,vitali22}.

Until the advent of Gaia, most of the analyses of the chemical composition of Sgr~dSph stars were focused on the nuclear region, because of its highest density of targets and intrinsic interest on the nucleus, and/or biased against metal-poor stars, as the  contamination by foreground MW stars of photometrically-selected samples was particularly strong on the blue side of the Sgr RGB, where metal-poor stars are located \citep{monaco2005ital,sbordone2007exotic,carretta2010m54+,mcwilliam2013chemistry,mucciarelli2017chemical}. \citet{minelli2023metallicity} discuss in detail how a target selection based on Gaia astrometry is effective in tackling the MW contamination, thus allowing unbiased sampling of stars spanning the entire metallicty range of the dwarf.
Indeed, a wealth of new studies to obtain a more complete view of the chemical composition of Sgr~dSph stars have been published in the latest years. 
\citet{hayes2020metallicity} and \citet{hasselquist2021apogee} discussed the abundances of 
$\alpha$-elements, Al, Ni and Ce derived 
from the near-infrared spectroscopic survey APOGEE for stars distributed in the main body and in the 
stream. \citet{hansen18}, \citet{sestito24} and \citet{ou25arx} investigated the chemical properties of some very 
metal-poor ([Fe/H]$< -$2 dex) Sgr stars, finding similar chemical properties between Sgr and MW stars of similar [Fe/H]. 
Also, \citet{vitali24arx} discussed the chemical composition of 111 Sgr giant stars located 
outside the nuclear region and observed with FLAMES-GIRAFFE. They measured chemical abundances of 13 elements 
(all of them analyzed also in this study) over a [Fe/H] range between --2.0 and --0.3 dex.

In this work, we provide a valuable addition to the studies mentioned above, by providing an extensive network 
of chemical abundances (21 elements, from Oxygen to Europium) for stars comprised in a large metallicity range 
(-2$\lesssim$[Fe/H]$\lesssim$-0.5) in the Sgr dSph main body.
We provide chemical abundances for some elements (i.e. Mn, Ni and Zn) which are crucial for understanding 
the chemical evolution of this galaxy and not extensively investigated in previous works.
The combination of such features is fundamental for understanding the chemical evolution pathway and the intrinsic 
properties of the stellar populations within this galaxy. 
In fact, the different families of elements probed in this study allow to characterize the enrichment by different 
stellar progenitors, from massive stars exploding as core-collapse supernovae (CC-SNe, producing most of the $\alpha$-elements,  
e.g. \citealt{romano2010quantifying}), to low- and intermediate-mass stars (which are important slow neutron-capture 
element producers at late times, e.g. \citealt{cescutti2022}) and Type Ia supernovae (SNe Ia, producing most of 
the Fe-peak elements, e.g. \citealt{kob_ln20,palla21}). 
In addition, this study provides results based on stellar spectra with a higher resolution than most other studies in this field.

The paper is organized as follows: Section~\ref{dataset} presents the spectroscopic  dataset analyzed in this study; 
Section~\ref{analysis} describes the methods used to infer the atmospheric parameters of the target stars, their radial 
velocities and chemical abundances; 
Sections~\ref{iron} and \ref{chems} describe and discuss the derived chemical abundances and the resulting abundance patterns of Sgr; 
Finally, Section~\ref{conclusions} summarizes the main results of this study.

\section{Dataset}
\label{dataset}
The dataset discussed here includes high-resolution spectra of 37 RGB stars, all of which belong to the Sgr galaxy. 
All the spectra were acquired with the high-resolution fiber-fed spectrograph UVES-FLAMES \citep{pasquini2002installation} 
mounted at the Very Large Telescope of ESO. For all the observations we adopted the UVES setup Red Arm 580 covering 
from 4800 to 6800 \AA\ and with a spectral resolution of 47,000. 

Twenty-three new targets were observed under the ESO program 105.20AH.001 (PI: Bellazzini) and 
selected on the blue side of the Sgr RGB in order to privilege metal-poor stars.
These 23 targets were selected following the same procedure described in \cite{minelli2023metallicity}, 
using the third data release (DR3) of the Gaia/ESA mission \citep{2016gaia,gaia21}. 
In particular, to minimize the contamination from foreground MW stars, that is especially strong on the blue side of the Sgr RGB, we selected only stars (1) with proper motions within $0.5$ mas/yr of the systemic proper motion of Sgr, as determined by \cite{gaia18helmi}, corresponding to $\rm \approx \pm 60$ km/s from the systemic motion of Sgr~dSph, and 
(2) with parallaxes ($\pi$) within $3.0\sigma$ from $\pi=0.0$~mas.

\begin{figure}[h!]
\includegraphics[scale=0.22]{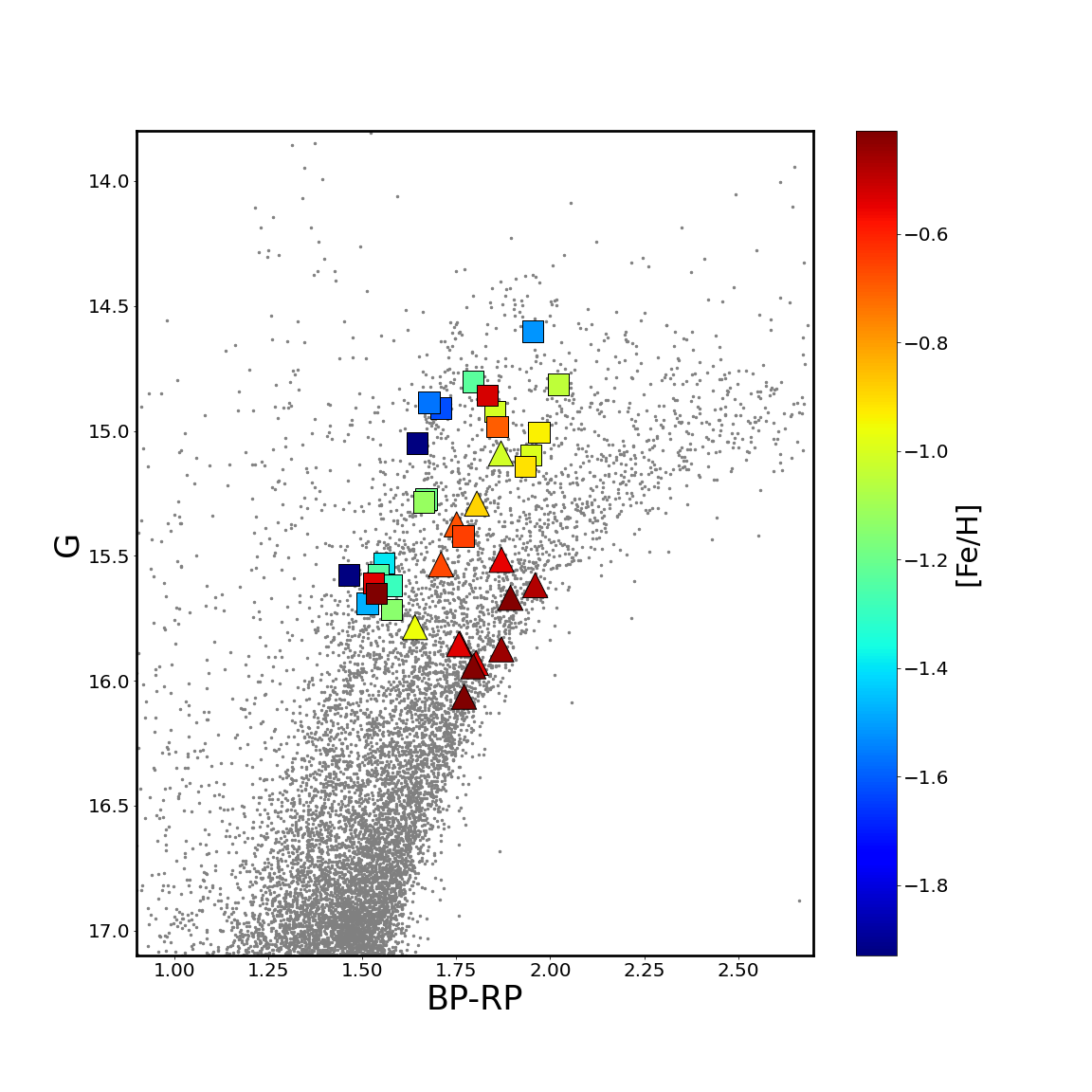}
\caption{ 
Position of the spectroscopic targets on the Gaia DR3 color-magnitude diagram of Sgr dSph 
(only Sgr stars outside the tidal radius of M54 are shown)
color-coded according to their [Fe/H].
Triangles are the targets already discussed in \citet{minelli2021homogeneous}, while squares are the new targets. 
}
\label{cmd}
\end{figure}

Among the bona-fide Sgr stars, the spectroscopic targets were selected in a radial region outside 
the tidal radius of M54  \citep[10.5'',][in order to avoid the contamination from cluster stars]{bellazzini2008nucleus} within 60' from Sgr center. 
Stars with G-band magnitude between 14.5 and 15.8 were considered, 
privileging stars in the bluer RGB of Sgr (see squared symbols in Fig.~\ref{cmd}).
We also exclude stars with bad photometric data, as traced by the Gaia quality parameter {\tt phot\_bp\_rp\_excess\_factor}, 
according to Eq. C.2 of \cite{lindegren2018gaia}.
Finally, to avoid contamination of the light collected by individual FLAMES fibers from (relatively) bright sources near 
our spectroscopic targets, we excluded stars of magnitude $G^*$ having a companion closer than $\rm 2.0\arcsec$ 
and brighter than $\rm G=G^*+1.0$.

Additionally, in order to sample the entire metallicity range covered by Sgr stars, we included in our sample 
fourteen stars, mainly belonging to the reddest RGB of the galaxy (the metal-rich, intermediate-age population, 
see the triangle symbols in Fig.~\ref{cmd}),  
using archival spectra obtained under the programs 71.B-0146 (PI: Bonifacio) and 081.D-286 (PI: Carretta).
These spectra have been previously analysed in \citet{monaco2005ital}, \citet{carretta2010m54+} 
and \citet{minelli2021homogeneous} and here we present a new analysis with a homogeneous approach with respect 
to the new targets.
The position of all the spectroscopic targets in the Gaia DR3 color-magnitude diagram of Sgr dSph is shown 
in Fig.~\ref{cmd}.

The spectra were reduced with the dedicated ESO pipeline\footnote{http://www.eso.org/sci/software/pipelines/} 
 UVES pipeline version 6.1.8 that performs bias subtraction, flat-fielding, wavelength calibration, spectral extraction, and order merging. 
The individual exposures were sky-subtracted using the average spectrum of two close sky regions observed in the same 
fiber configuration. 
The final signal-to-noise (S/N) ratio per pixel at 6300 \AA\ ranges from 20 to 40. 


\section{Analysis}
\label{analysis}

\subsection{Atmospheric parameters}
\label{subsect:atmospheric_param}
\begin{figure}[h!]
\includegraphics[scale=0.56]{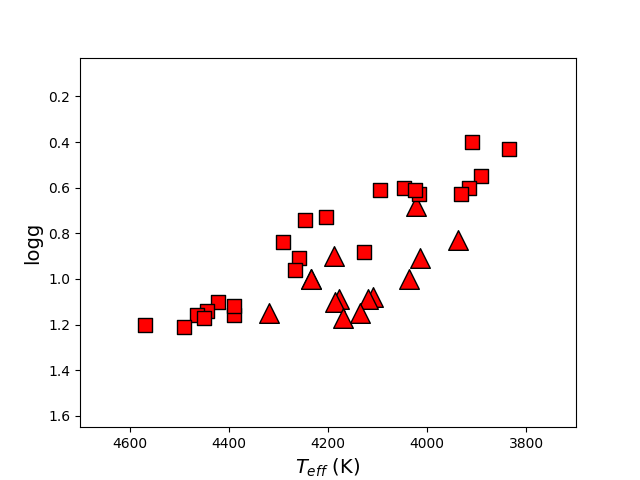}
\caption{ Kiel diagram showing the run of \gr\ as a function of \teff\ 
for the Sgr spectroscopic targets (same symbols as Fig.~\ref{cmd}).
}
\label{kiel}
\end{figure}

The effective temperature (\teff), the surface gravity (\gr)  and the microturbulence velocity (\vt) were 
obtained using the Gaia DR3 photometry \citep{gaia21}. 
We first derived the dereddened Gaia magnitudes using the relation by \cite{riello2021gaia} and adopting 
the color excess $E(B-V)=0.15$ \citep{layden2000photometry}. 
The effective temperatures were obtained using the $(BP-RP)_0$--\teff\ transformation by \cite{mucciarelli2021exploiting}. 
At the first step, \teff\ were computed assuming a metallicity of [Fe/H]=--1.5 dex for all the stars, then \teff\ 
were re-calculated adopting the appropriate metallicity of each target according to the results of the chemical analysis.
Surface gravities were derived adopting the above \teff\ , a distance of $D = 26.0 \pm 1.3$ kpc \citep{monaco2004distance}, 
a stellar mass of 0.8 $M_{\odot}$ and the G-band bolometric corrections obtained following the prescriptions by \cite{andrae2018gaia}.  Fig.~\ref{kiel} shows the position of the spectroscopic targets in the 
Kiel \teff-\gr\ diagram.
The microturbulence velocities were derived using the relation by \cite{mucciarelli2020facing} in order to 
avoid possible biases against the weak lines that can affect low S/N ratio spectra and therefore the derived values 
of \vt\ .
    
Uncertainties in \teff\ are dominated by the uncertainty in the adopted color-\teff\ transformation 
($\approx 80$ K, see \citealp{mucciarelli2021exploiting}), 
while the contribution by photometry and reddening errors is less than 10 K. 
Uncertainties in \gr\  are of about 0.1, including the contribution of errors in \teff\, 
adopted distance and stellar mass. 
The typical error on \vt\ is 0.15 km/s, calculated by adding in quadrature the error on the \gr\  
with the error associated to the relation used. 

\subsection{Radial velocities}
Heliocentric radial velocities (RVs) were measured with DAOSPEC \citep{stetson2008daospec} using a list of unblended lines selected 
for the chemical analysis (see Section~\ref{lines}). 
This code automatically finds the centroid of spectral lines by Gaussian fitting and derives the final RV as the average value
from the wavelength shifts of the measured lines. 
The associated uncertainties were computed as the standard deviation divided by the root mean square of the number of used lines.
As presented in Fig. \ref{fig:rv_fe}, the stars in our sample have RVs ranging from +119.4 to +158.9 km/s, consistent with the RV distribution of the Sgr stars 
(see, e.g., \citealt{minelli2023metallicity}) and confirming their membership to Sgr.
The associated errors are generally of $\rm 0.1 \ km/s$.
The radial velocities are listed in Table~A.1. 

\subsection{Line selection and spectral analysis}
\label{lines}
We derived the abundances of 21 species, namely O, Na, Mg, Al, Si, Ca, Sc, Ti, V, Cr, Mn, Fe, Co, Ni, Cu, Zn, 
Zr, Ba, La, Nd and Eu\footnote{All the abundances will be available in an electronic format.}. All the abundances were derived with our own code SALVADOR 
performing a $\chi^2$-minimization 
between observed line profiles and grids of synthetic spectra. The latter were calculated with the code SYNTHE \citep{kurucz2005atlas12} 
and adopting model atmospheres calculated with the code ATLAS9 \citep{castelli2003modelling}. 
We adopted the atomic and molecular linelist available 
in the R. L. Kurucz\footnote{http://kurucz.harvard.edu/molecules.html} 
and F. Castelli\footnote{https://wwwuser.oats.inaf.it/fiorella.castelli/linelists.html} websites, updated for some 
specific transitions with more recent or more accurate atomic data. 
In particular, we adopted new gf-values for Sc \citep{lawler19sc}, Ti \citep{lawler13ti}, 
V \citep{lawler14v}, Ni \citep{wood14ni}, Zn \citep{roederer12zn}. 
Hyperfine/isotopic splittings are included for all the Sc, V, Mn, Co, Cu, Ba, La and Eu transitions.
For each star, we selected a list of metallic transitions predicted to be unblended according 
to the observed spectral resolution and to the metallicity and stellar parameters of the target, 
following the iterative scheme described in \citet{mucciarelli23a}.
We privileged transitions with laboratory oscillator strengths. 
For Na abundances only, we applied corrections for Non-Local Thermal Equilibrium (NLTE) 
using the corrections grid by \citet{lind2011non}. 

The total uncertainty in each abundance ratio was computed by adding in quadrature 
the two main sources of errors, namely the error in the line fitting procedure and 
those arising from the stellar parameters. The uncertainty in the line-fitting 
procedure was computed resorting to Montecarlo simulations. For each star, we created 
a set of 500 noisy synthetic spectra with a Poissonian noise that reproduces the observed S/N, 
and the line-fitting procedure was repeated. The dispersion of the abundance distribution obtained from these 
simulated spectra was assumed as the 1$\sigma$ uncertainty. 
The abundance errors arising from the parameters were obtained by repeating the chemical analysis 
varying each time one parameter by its 1$\sigma$ uncertainty.
We refer the reader to \citet{mucciarelli13foster} 
and \citet{minelli2021homogeneous} for additional details about the uncertainties estimate.

\begin{figure}[h!]
\includegraphics[scale=0.57]{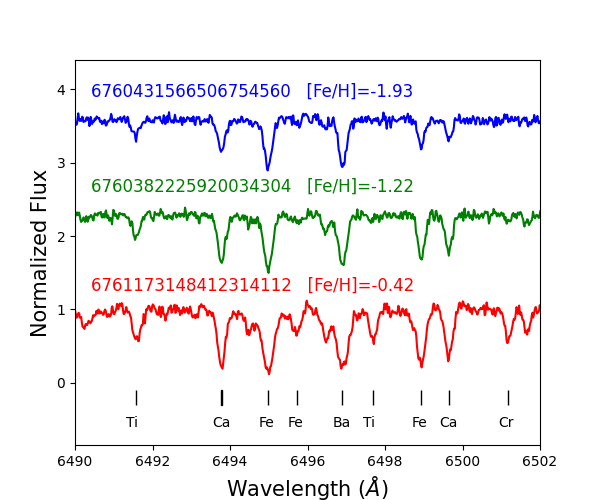}
\caption{ 
Examples of spectra for three target stars with different [Fe/H] and marked 
some lines of interest.}
\label{spectra}
\end{figure}


\section{Iron content}
\label{iron}

The sample of Sgr stars covers a range of [Fe/H] between $-$1.93 and $-$0.41 dex. 
 Figure ~\ref{cmd} shows the position of the targets in the Gaia DR3 color-magnitude diagram 
color-coded according to their [Fe/H].
While almost all the stars of this sample agree well with the color-metallicity relation 
usually observed in Sgr \citep{bellazzini2008nucleus,alfaro2019,vitali22}, 
3 of them have much bluer colors than the bulk of the Sgr stars of the same metallicity ([Fe/H]$\sim -$0.7/$-$0.6 dex)
that lie on the red side of the RGB (see Fig.~\ref{cmd_fe}).

While metal-rich stars on the red side of the RGB should have old ages ($\sim$8--10 Gyr), 
stars with [Fe/H]$\sim -$0.7/$-$0.6 dex located on the blue side of the RGB 
should be significantly younger. 
 Figure~\ref{cmd_fe} shows some BaSTI-IAC isochrones \citep{pietr21} 
with [Fe/H]$= -$0.6 and $-$0.4 dex, solar-scaled chemical mixture and different 
ages. It is found that a metal-rich population with age $\sim$1--2 Gyr perfectly overlaps 
the blue side of the Sgr RGB, where are also located three metal-rich Sgr stars.

We can envisage two possible scenarios to explain the existence of these stars.
The first scenario is that Sgr hosts a metal-rich ([Fe/H]$\sim$--0.7/--0.6 dex) 
and young ($\sim$1--2 Gyr) population. This population should have had a distinct chemical enrichment path 
with respect to that of the dominant Sgr population of similar [Fe/H], with ages of 8--10 Gyr. 
This scenario implies the existence of a secondary branch of the Sgr age-metallicity relation, 
suggesting a higher level of complexity in the Sgr history than previously thought, 
for instance accretion of satellites \citep[see e.g.][]{davies24} or
late gas infall/accretion triggering star formation \citep[see e.g.][]{spitoni2022,palla2024}. 
 Note that the age of this metal-rich population corresponds to the epoch when 
the gas was totally stripped \citep[see e.g.][]{tepper2018}.
Such a metal-rich, young population should have a main sequence that well overlaps the extended 
blue plume of Sgr \citep{bellazzini06} and probably it could be its dominant component. 
Alternatively, these stars are not genuine young stars but they are the product of mass-transfer 
in binary systems \citep[see e.g.][]{mccrea64}. However, 
no sign of photometric or astrometric anomaly that can be connected to stellar multiplicity \citep[e.g., the RUWE parameter][]{} is apparent in the Gaia DR3 catalog.

In the following figures showing the measured abundance ratios we highlight the position of these three targets.
We have not found significant differences in the chemical composition of metal-rich stars located on the blue and red RGB, 
likely expected if these groups of stars have reached the same [Fe/H] in significantly different timescales.
However, the small size of our sample and the relatively high abundance uncertainties
highlight the need for further investigation (as supported also 
by observations in the Galaxy and the solar vicinity, where uncertainties reach much lower levels). 
The presence of these blue, metal-rich stars surely deserves a deeper investigation.

\begin{figure}[h!]
\includegraphics[width=\hsize]{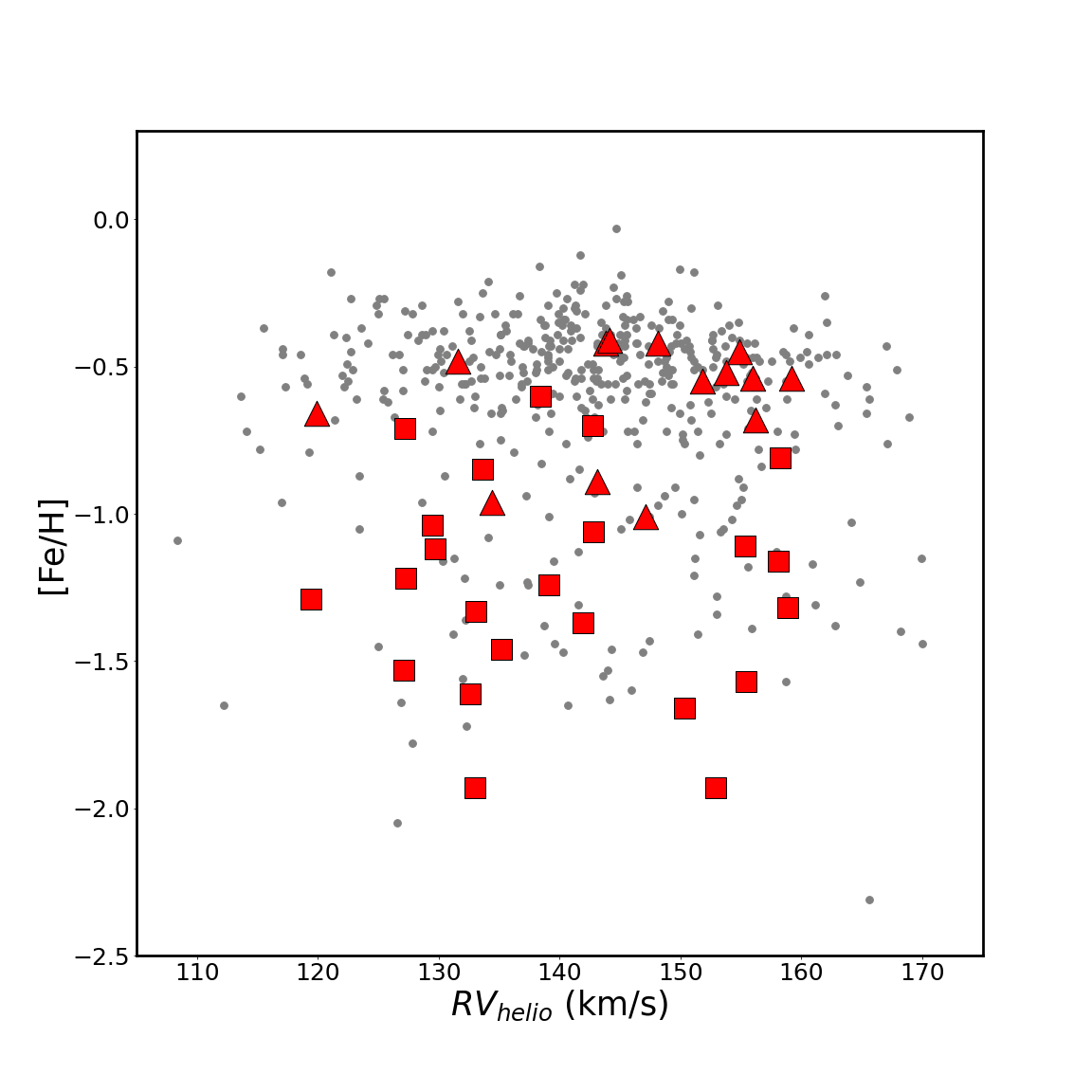}
\caption{Behavior of [Fe/H] as a function of RV for the spectroscopic dataset discussed here 
(same symbols of Fig.~\ref{cmd}) in comparison with the sample of Sgr stars by \citet{minelli2023metallicity}.}
\label{fig:rv_fe}
\end{figure}

\begin{figure}[h!]
\includegraphics[scale=0.22]{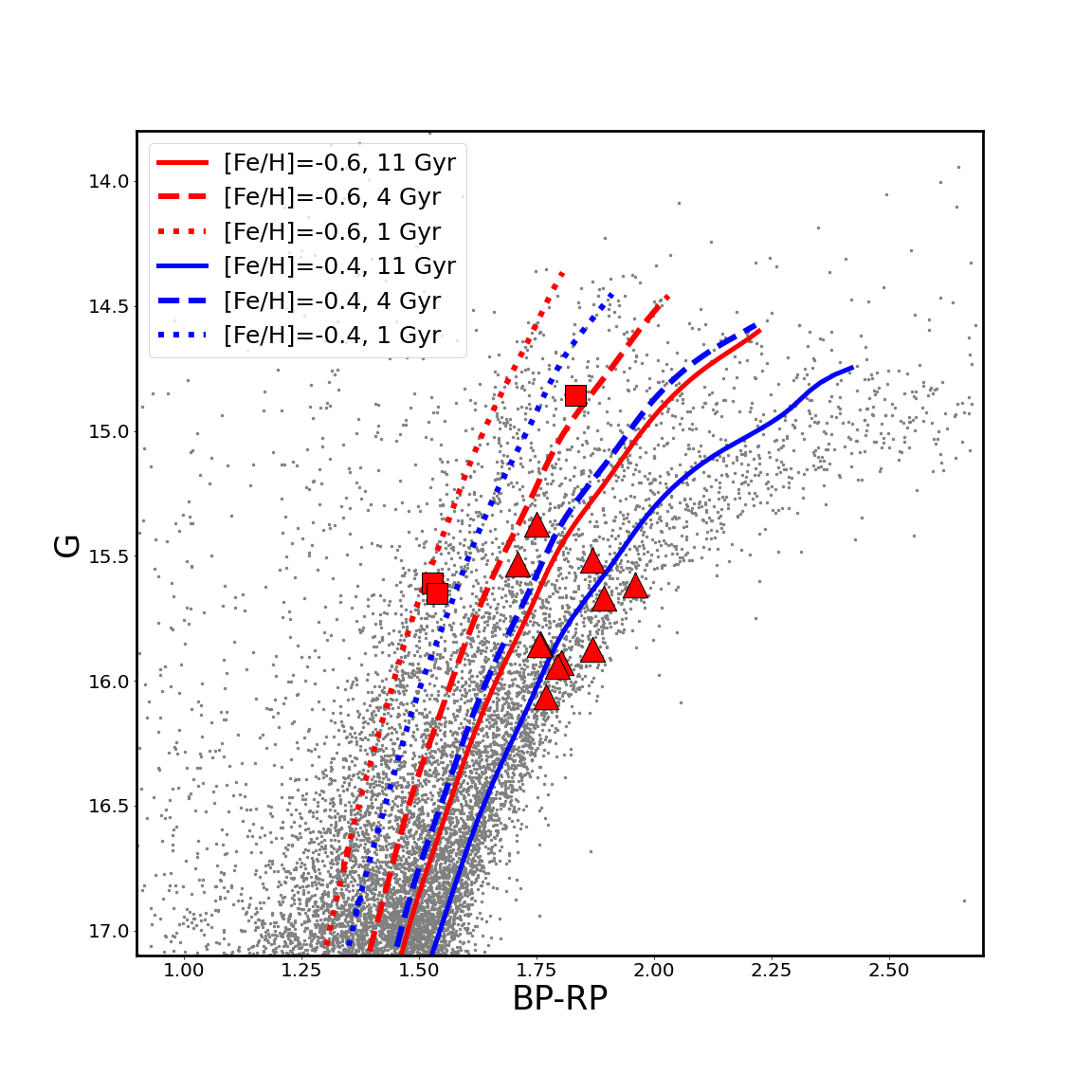}
\caption{Position in the color-magnitude diagram of the Sgr targets with [Fe/H]$>$--0.7 dex, superimposed to 
some BaSTI-IAC isochrones \citep{pietr21} with [Fe/H]=--0.6 and --0.4 dex, solar-scaled chemical mixture, red and blue curves respectively, and ages of 1, 4 and 11 Gyr.}
\label{cmd_fe}
\end{figure}


\section{The chemical composition of Sgr}
\label{chems}

Here we present a complete screening of the chemical composition\footnote{All the abundance ratios and their uncertainties 
are available in electronic form.} of Sgr stars 
over a large [Fe/H] range. The metal-poor stars of the sample were selected 
to avoid contamination by the globular cluster M54.
 In fact, the presence of the latter stars in spectroscopic samples obtained in Sgr stellar nucleus 
does not allow to unveil the real chemistry of this galaxy between [Fe/H]$\simeq -$2 and $-$1 dex. 
Therefore, our selection criteria allow to reveal the proper evolution of the progenitor of Sgr dSph.

In Figures \ref{oddz}--\ref{ncap}, the abundance ratios of the Sgr targets are compared with the sample of 
Galactic GCs analyzed by \citet{mucciarelli23b} adopting the same assumptions of this analysis 
(i.e. model atmospheres, atomic data for the used transitions, software for the analysis, etc.), thus providing the benchmarck for a fully homogeneous comparison with the measures presented here. 
 Only for the four elements involved in the anticorrelations phenomenon \citep[namely O, Na, Mg, Al; see, e.g.,][]{gratton2004}
we considered for each cluster only first population stars. In fact, these stars have abundances that well 
match with those of MW field stars of similar metallicity, at variance with the second generation stars 
whose abundances of these four elements reflect the internal chemical evolution of the clusters. 
For all the other elements we don't distinguish between first and second generation stars. 
In addition, we show also the MW field stars available in the SAGA database \citep{saga} 
that provide a representative (even if not homogeneous) reference for the overall chemical patterns 
over a large range of [Fe/H].  Despite possible systematics in the literature sample, it is useful to display the
overall trends in the MW based on a large number of stars.
 However, the comparison between the chemical properties of Sgr and MW is based on this reference sample.
Sgr stars exhibit, for most of the elements, trends with [Fe/H] that stand out from those of the MW stars. 
In the following, we provide the details of these trends for different families of elements 
(odd-Z, $\alpha$, Fe-peak and neutron-capture elements).

\subsection{Odd-Z elements (Na and Al)}

Sodium and Aluminium are mostly produced in massive stars during hydrostatic C (and Ne, for Al) burning \citep{woosley1995}. 
The production of these elements largely depends on stellar metallicity, as it needs for neutron seeds that trace back 
to the nitrogen produced CNO cycle (e.g. \citealt{kobayashi06}). 
Moreover, a small (but non-negligible) contribution comes from lower-mass asymptotic giant branch (AGB) stars 
\citep[see][and references therein]{smiljanic2016}.

\begin{figure*}
\centering
\includegraphics[scale=0.22]{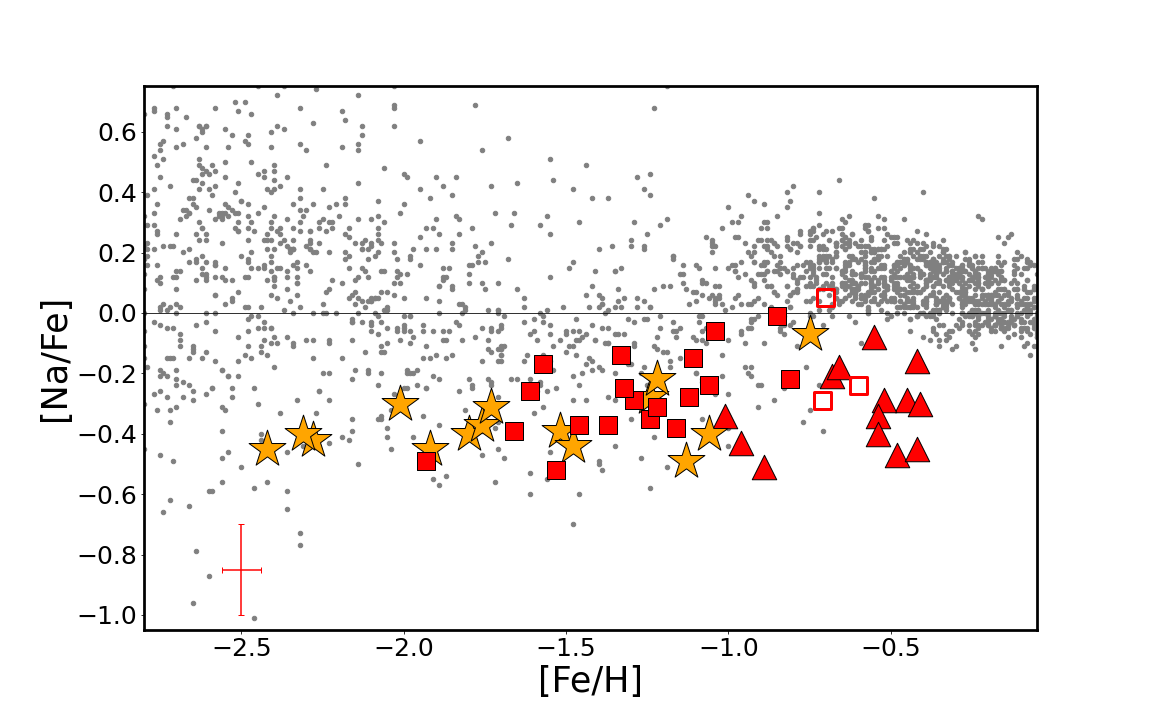}
\includegraphics[scale=0.22]{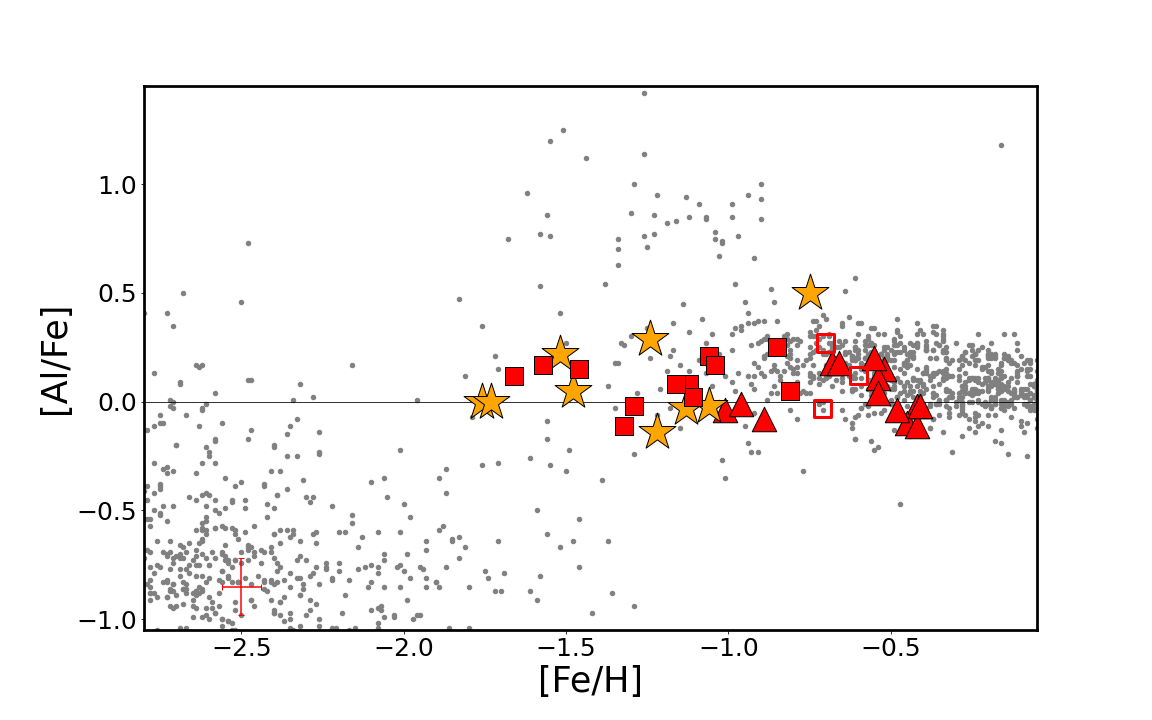}
\caption{Behavior of [Na/Fe] and [Al/Fe] as a function of [Fe/H] for the Sgr spectroscopic targets of this study 
(same symbols of Fig.~\ref{cmd}), in comparison with the abundances of MW GCs analyzed by \citet[][orange star symbols]{mucciarelli23b} 
and with the MW field stars from the SAGA database \citep[][grey circles]{{saga}}. 
Empty red squares are the metal-rich Sgr targets belonging to the blue side of Sgr RGB 
(see Section~\ref{iron}).
The typical errorbar of the abundance ratios 
measured for Sgr stars is shown in the left-bottom corner.}
\label{oddz}
\end{figure*}

The observed patterns for Na and Al are shown in Fig. \ref{oddz}.
For Na, the observed stars in Sgr dSph show a mildly increasing trend with metallicity. 
For [Fe/H]$< -$1 dex, [Na/Fe] in Sgr stars well match those traced by first-generation MW GCs stars and agrees with 
the theoretical expectations as more metallic massive stars exhibit larger Na yields. Such a trend is instead only 
partly seen in MW field stars, where the average trend shows an upturn only for [Fe/H]$\simeq -$1.5 dex and generally 
higher [Na/Fe] values \citep[see also][their figure~8, left panel]{smiljanic2016}. 
This offset between Sgr/MW GC stars and the MW field stars could be explained 
by the different NLTE corrections  (being the Na abundances significantly affected by the 
departures from LTE also at the metallicity of our targets), with the first sample including the corrections by \citet{lind2011non} while 
the second one including different (or lacking) NLTE corrections.

At [Fe/H]$\simeq -$1 dex, [Na/Fe] in Sgr remain low, while in MW field stars and in 47 Tucanae (the most metal-rich GC of our reference sample) 
there is an increase of this abundance ratio. 
This can be attributed to SNe Ia pollution, that in Sgr starts to become effective at lower metallicity than in the MW. 
However, in the metallicity range covered by our sample stars, we observe a large scatter in [Na/Fe], which cannot be explained in the 
light of typical abundance uncertainties. 
This scatter might reflect multiple sites of Na 
production, coupled to inhomogeneous chemical enrichment. Similarly large [Na/Fe] scatters have been
observed in other MW satellites, i.e. the Large Magellanic Cloud 
\citep[LMC][]{pompeia08,vds13,minelli2021homogeneous}, the Small Magellanic Cloud \citep[SMC][]{mucciarelli23a}
and Fornax \citep{letarte2010}.

For what concerns Al, the trend at lower metallicities ([Fe/H]$< -$1 dex) remains quite similar to what is seen in the MW 
(especially for GCs, as field stars are lacking in the range $-2<$[Fe/H]/dex$<-1$). However, we can note that no 
increasing [Al/Fe] trend is observed or, at least, it is hidden in the scatterplot. For larger metallicities, 
we instead see an evident downturn that detaches the Al trend observed in Sgr from the one in the MW 
\citep[see, however,][their figure~8, right panel, showing a prominent downturn in MW Gaia-ESO data]{smiljanic2016}. 
Both the non-increasing trend at low metallicities and the decreasing one for the most metallic Sgr \
stars can be easily explained in the context of the time-delay model 
\citep[e.g.][see also Sect.~\ref{sec:alphas}]{matteucci90,matteucci2012,matteucci2021},
in which the relative contribution from SNe Ia to the ISM pollution is more prominent at lower metallicities 
in dwarf galaxies (such as Sgr) relative to actively star-forming galaxies, such as the MW.  
Our results for Na and Al are compatible with those obtained by \citet{vitali24arx} in a similar 
[Fe/H] range. On the other hand, \citet{hasselquist2021apogee} found [Al/Fe] around $\sim$--0.5 dex 
at any metallicity  (see Fig.~\ref{appendix:hass}). The origin of this discrepancy is not clear and 
maybe it could be due to some  relevant systematics between optical and near-infrared Al lines 
 (these systematics among different diagnostics can also affect other elements).

\subsection{$\alpha$-elements}
\label{sec:alphas}

The so-called $\alpha$-elements are produced mainly in massive stars and are usually grouped in elements produced 
during hydrostatic and explosive burnings. 
O and Mg belong to the first class and they are almost completely produced in massive stars \citep[see e.g.][]{romano2010quantifying,palla2022}, 
while Si, Ca and Ti are produced through explosive nucleosynthesis with a minor but not negligible contribution 
by SNe Ia \citep[see.e.g.][]{seitenzahl2017,kobayashi2020origin}.
Since from some seminal papers \citep{tinsley79,matteucci_greggio86,matteucci90}, the [$\alpha$/Fe] abundance ratios 
have been recognized as a diagnostic sensitive to the star formation efficiency of the system, owing to the different 
time scales of enrichment by CC-SNe and SNe~Ia.\\ 
All the [$\alpha$/Fe] ratios in our sample of Sgr stars show decreasing trends by increasing [Fe/H]. 
The most metal-poor Sgr stars have enhanced [$\alpha$/Fe] ratios comparable with those measured 
in the GCs of the reference sample. Both \citet{sestito24} and \citet{ou25arx} found similar 
enhanced values for Sgr stars with [Fe/H]$<$--2.5 dex.

The metallicity of the $\alpha$-knee (corresponding to the metallicity where the contribution of SNe Ia 
starts to lower significantly the [$\alpha$/Fe] ratios) is around [Fe/H]$\sim -$1.5/$-$1.3 dex, 
as well visible in the case of oxygen that exhibits the cleanest trend.
\citet{vitali24arx} attempt to constrain the position of the knee using data from various studies and propose 
a value of [Fe/H]$= -$1.05 dex, while highlighting some weaknesses of this value (statistics, data quality, and heterogeneity in analyses). 
Our [Mg/Fe], [Si/Fe] and [Ca/Fe] abundance ratios well match with those obtained by \citet{hasselquist2021apogee}, 
with the only discrepancy in the [Mg/Fe] of the metal-poor stars, where APOGEE provides [Mg/Fe] values 
lower by $\sim$0.2 dex than our results (see Fig.~\ref{appendix:hass}). As for Al, also for 
these elements possible systematics between optical and near-infrared lines can affect the abundances.

\begin{figure*}
    \centering
    \includegraphics[scale=0.22]{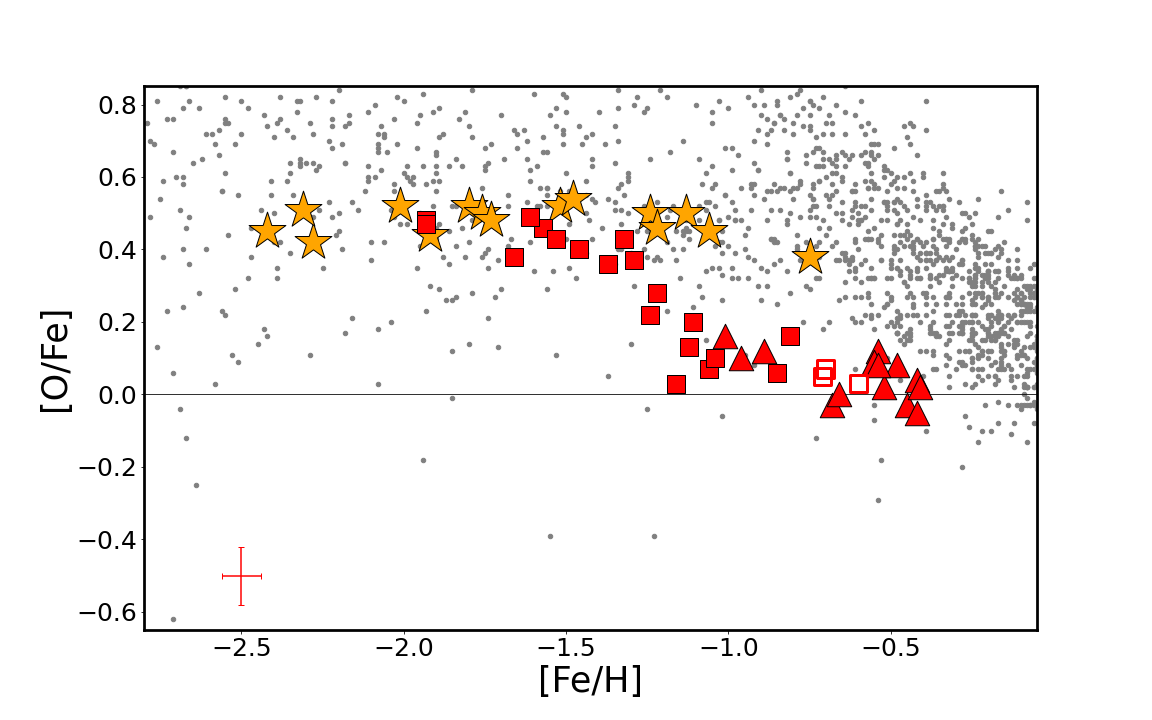}
    \includegraphics[scale=0.22]{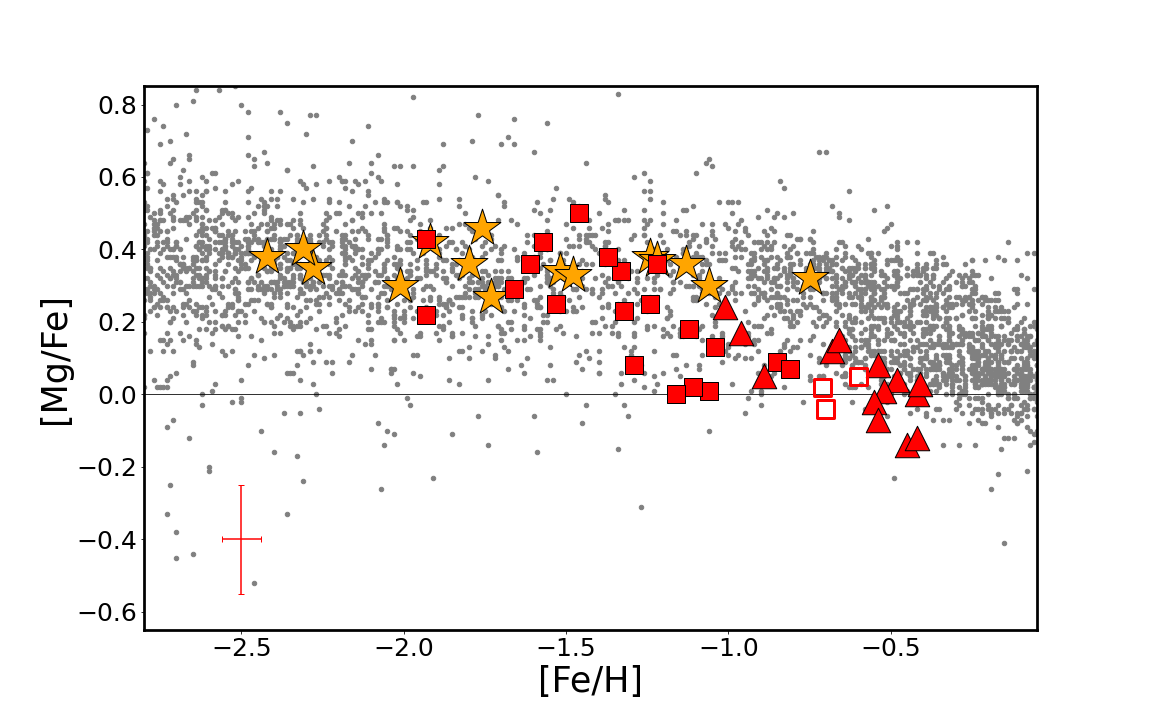}
    \includegraphics[scale=0.22]{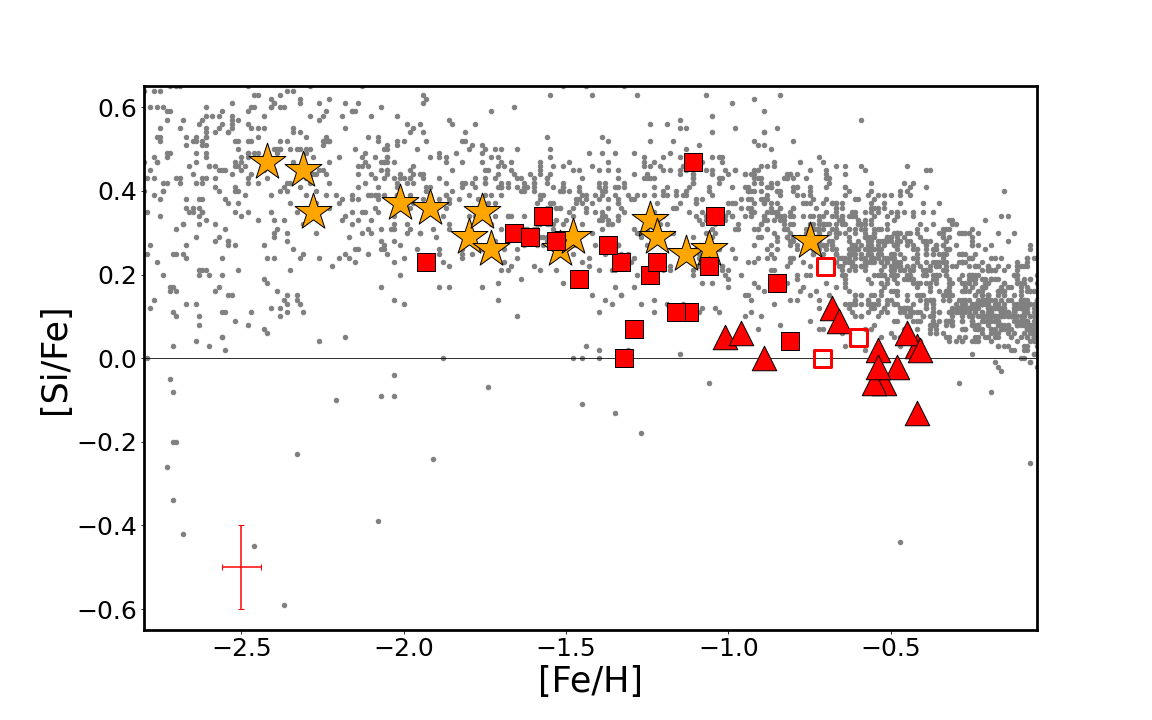}
    \includegraphics[scale=0.22]{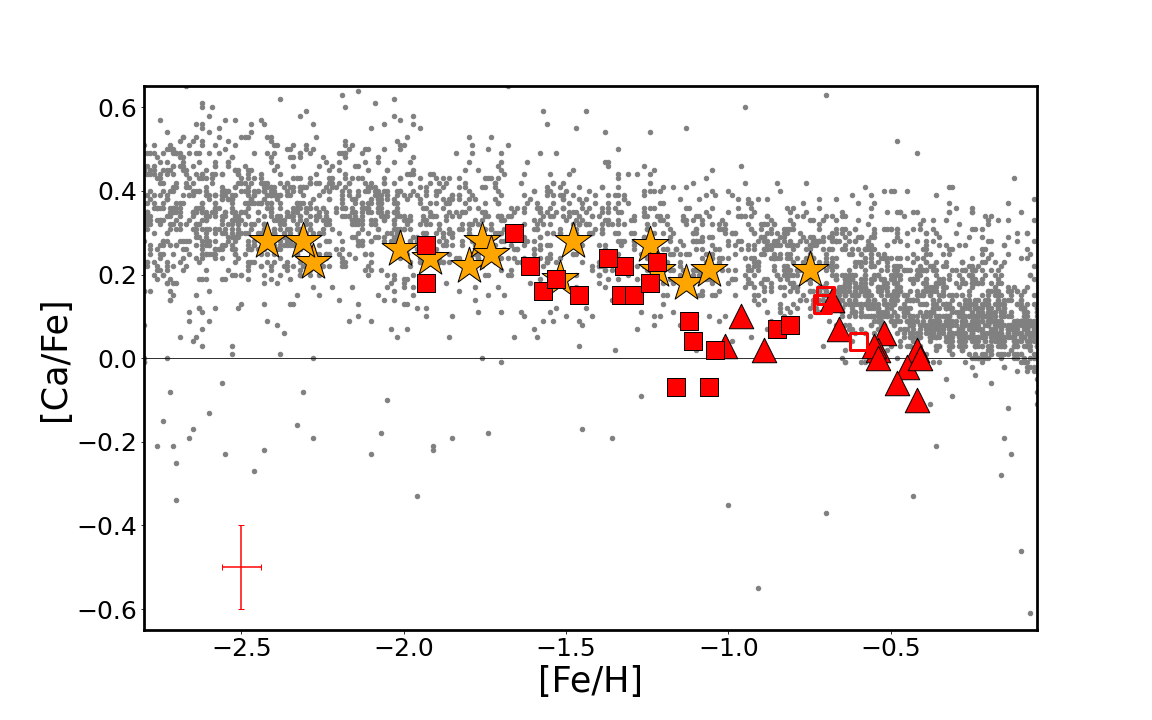}
    \includegraphics[scale=0.22]{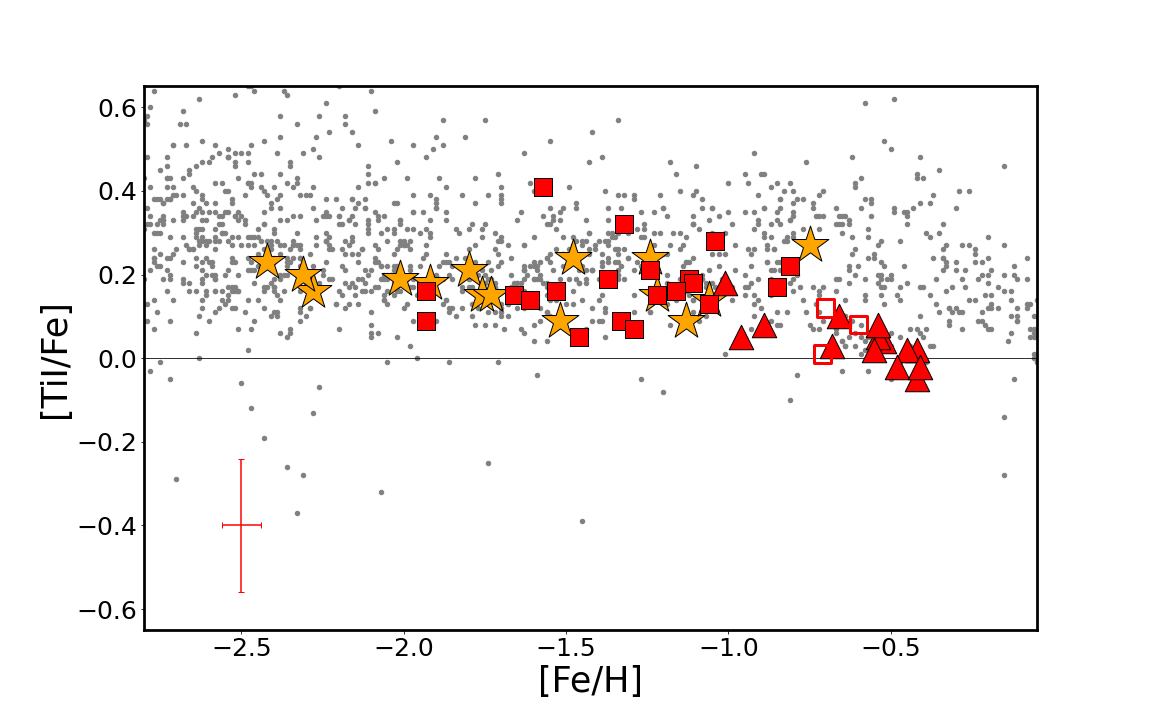}
    \caption{$\alpha$-elements (O, Mg, Si, Ca and Ti) abundance ratios as a function of [Fe/H], same symbols of Fig~\ref{oddz}.}
    \label{alfa}
\end{figure*}

The behavior that we found is consistent with the lower star formation 
efficiency of Sgr \citep{mucciarelli2017chemical} compared to the MW galaxy. 
Indeed, looking at the different panels of Figure \ref{alfa}, both MW GC and field stars 
exhibit a plateau in [$\alpha$/Fe] ratios extending up to higher metallicities ([Fe/H]$\sim -$1 dex). 
This is indicative of higher efficiency of star formation, as it is explained by a larger number of prompt sources 
(namely, CC-SNe) polluting the ISM before SNe Ia start to be effective.

\begin{figure}[h!]
\includegraphics[width=\hsize]{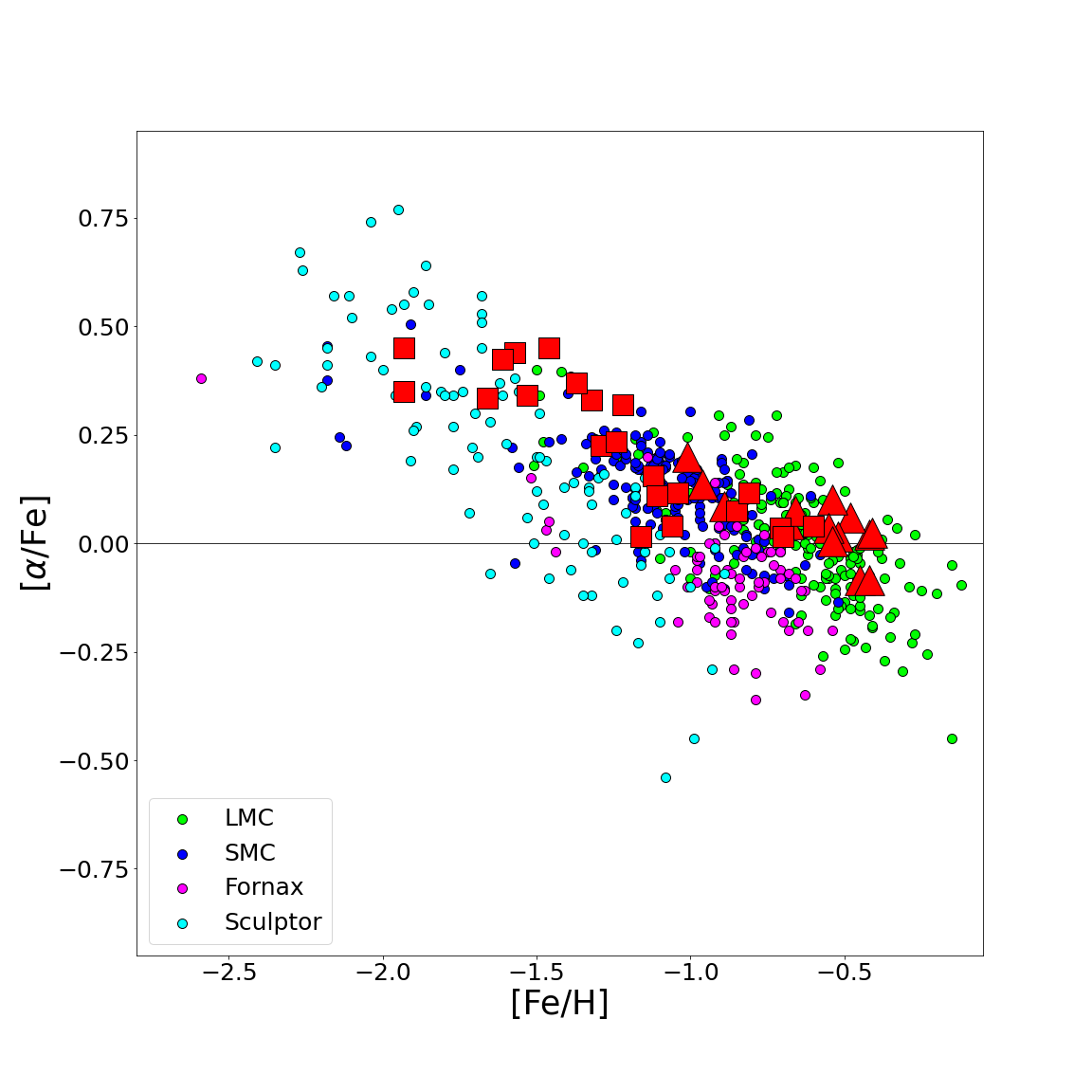}
\caption{Behavior of the average hydrostatic [$\alpha$/Fe] as a function of [Fe/H] for the Sgr stars (same symbols of 
Fig.~\ref{alfa}) in comparison with stars of the LMC \citep{vds13,lapenna12}, 
the SMC\citep{mucciarelli23a}, Fornax \citep{letarte2010} and Sculptor \citep{hill19}. 
The [$\alpha$/Fe] is computed by averaging [O/Fe] and [Mg/Fe].}
\label{alfa2}
\end{figure}

Finally, Fig.~\ref{alfa2} compares the average value of the hydrostatic $\alpha$-elements of the Sgr stars 
with the abundance patterns observed in other dwarf galaxies in the Local Group, i.e. LMC 
\citep{lapenna12,vds13}, SMC \citep{mucciarelli23a}, Fornax \citep{letarte2010} 
and Sculptor \citep{hill19}. The trend defined by Sgr stars well matches that drawn by the 
Magellanic Clouds stars, especially for the most metal-rich range and in the flat metal-poor branch, that is absent in lower mass dwarfs. This suggests a similar 
chemical enrichment history between Sgr and these irregular galaxies, with the match in the range [Fe/H]$\geq-0.8$ suggesting that the mass of the Sgr progenitor may have been
comparable with that of the LMC, as already pointed by other studies \citep{deboer15x, gibbons2017tail,johnson2020diffuse,minelli2021homogeneous}. 
On the other hand, the $\alpha$-sequence of Sgr stars clearly separates from that of dwarf spheroidal galaxies 
like Sculptor or Fornax, indicating a more intense chemical enrichment compared to these galaxies.

\subsection{Iron-peak elements}
\begin{figure*}[!h]
    \centering
    \includegraphics[scale=0.22]{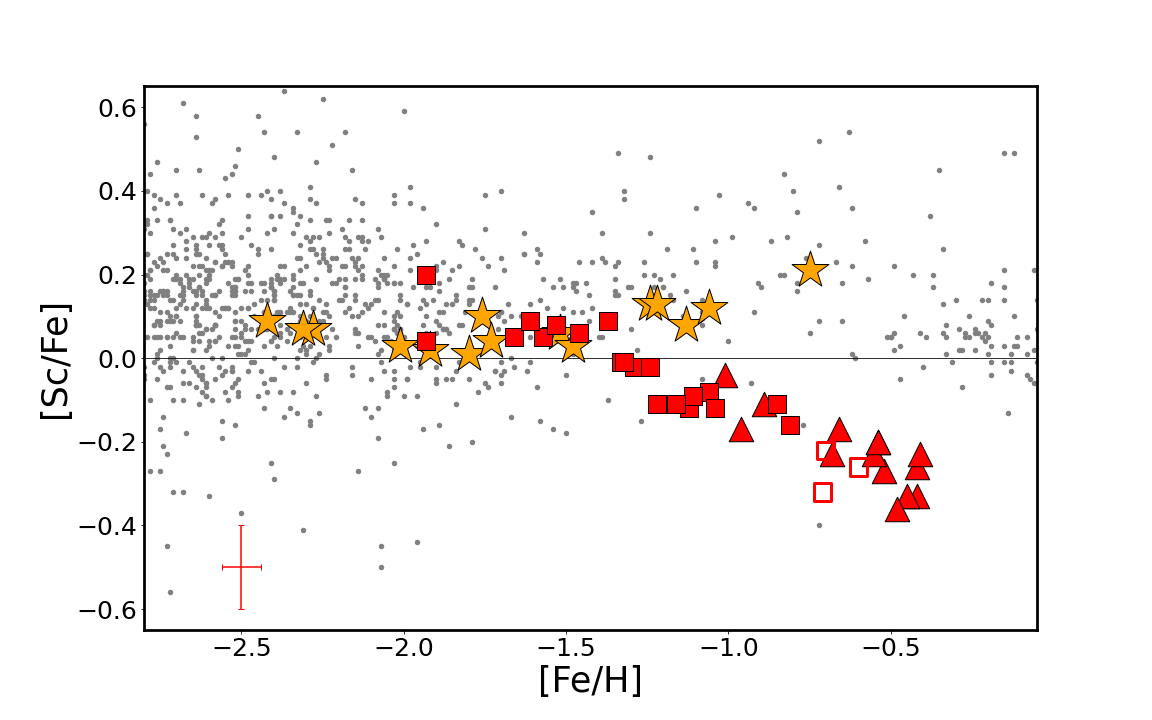}
    \includegraphics[scale=0.22]{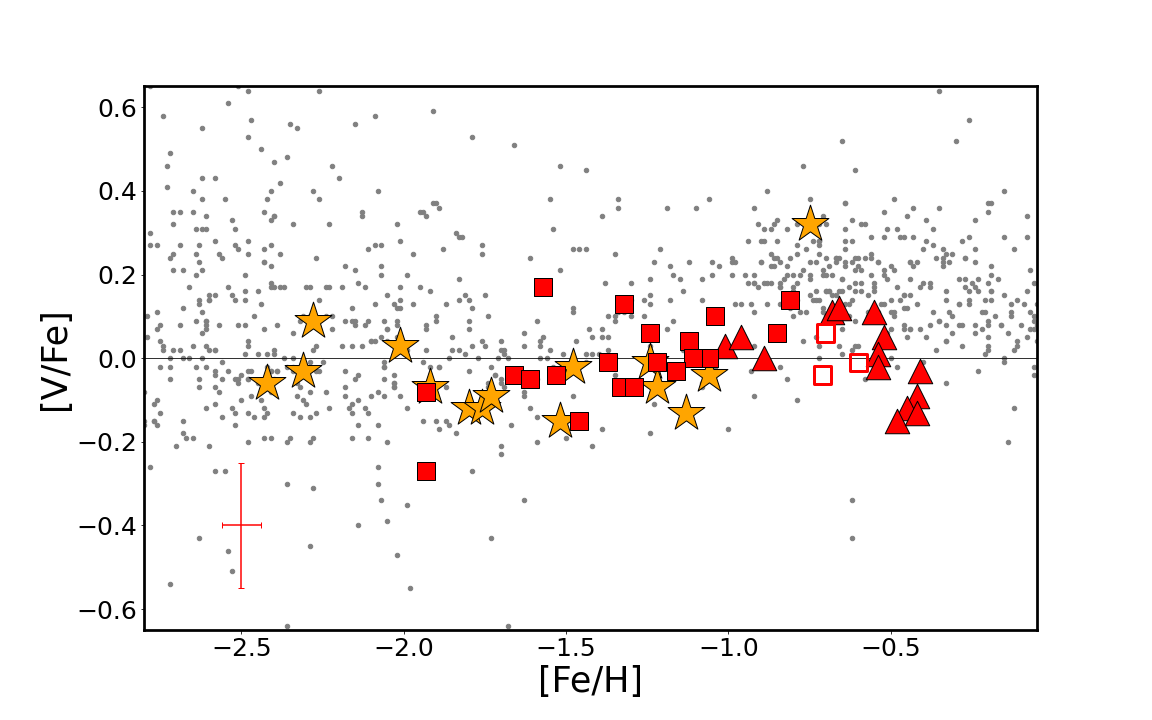}
    \includegraphics[scale=0.22]{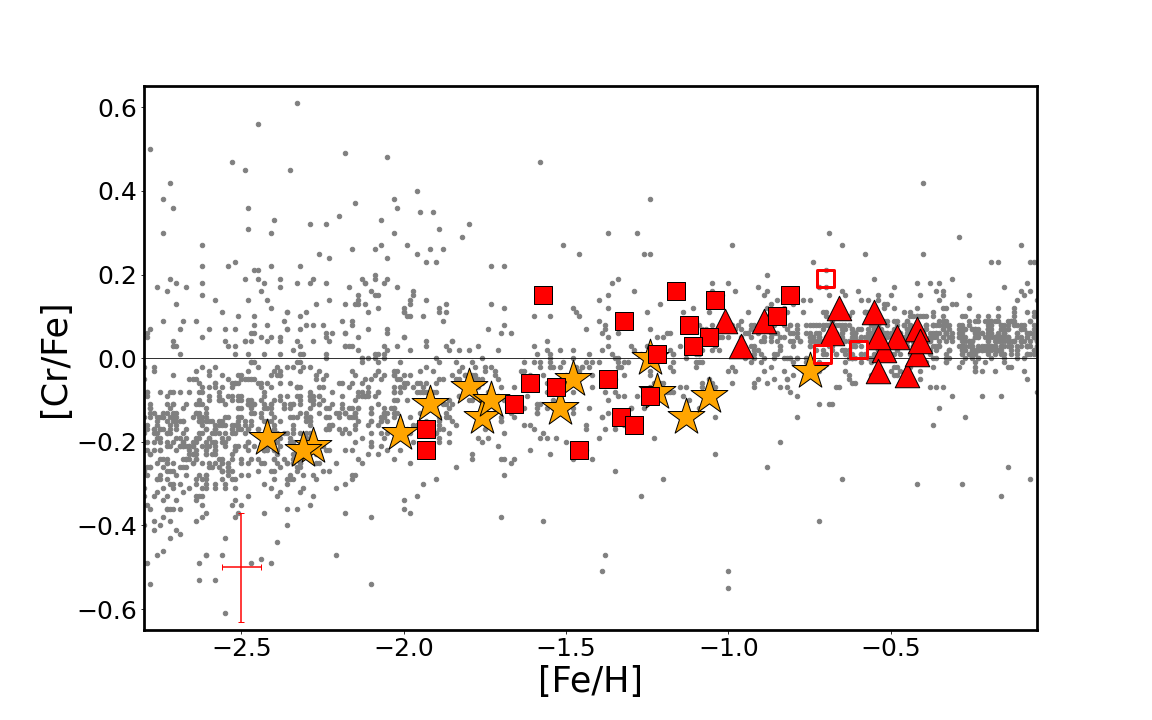}
    \includegraphics[scale=0.22]{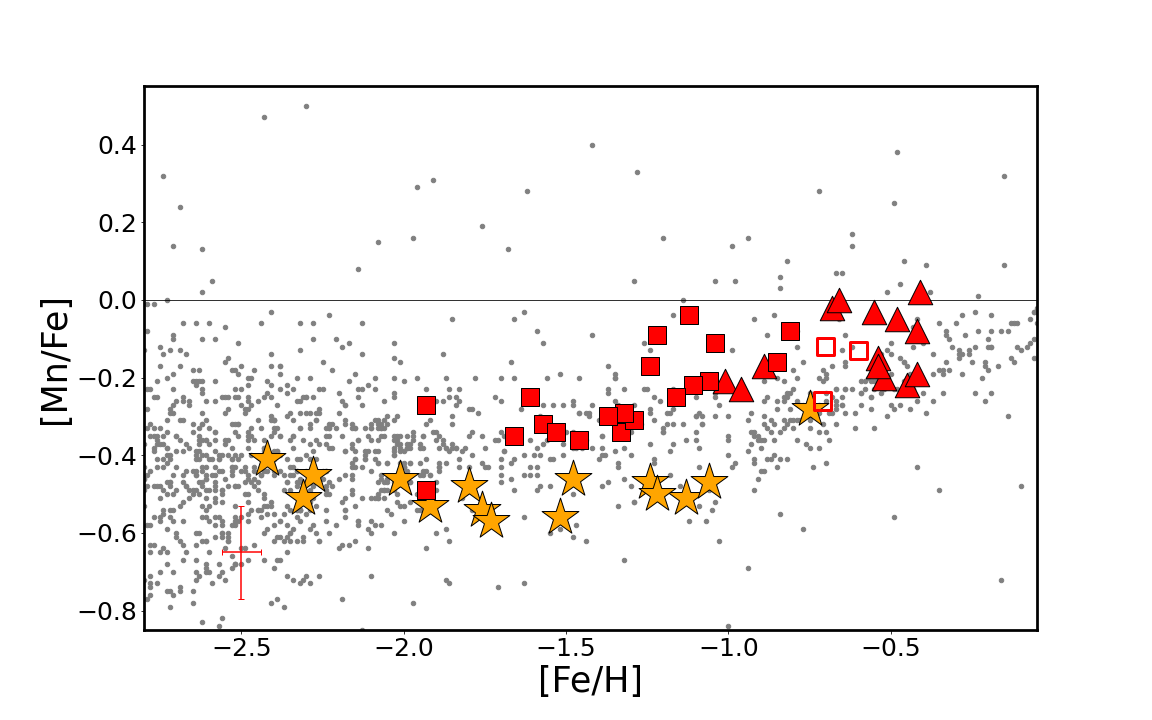}
    \includegraphics[scale=0.22]{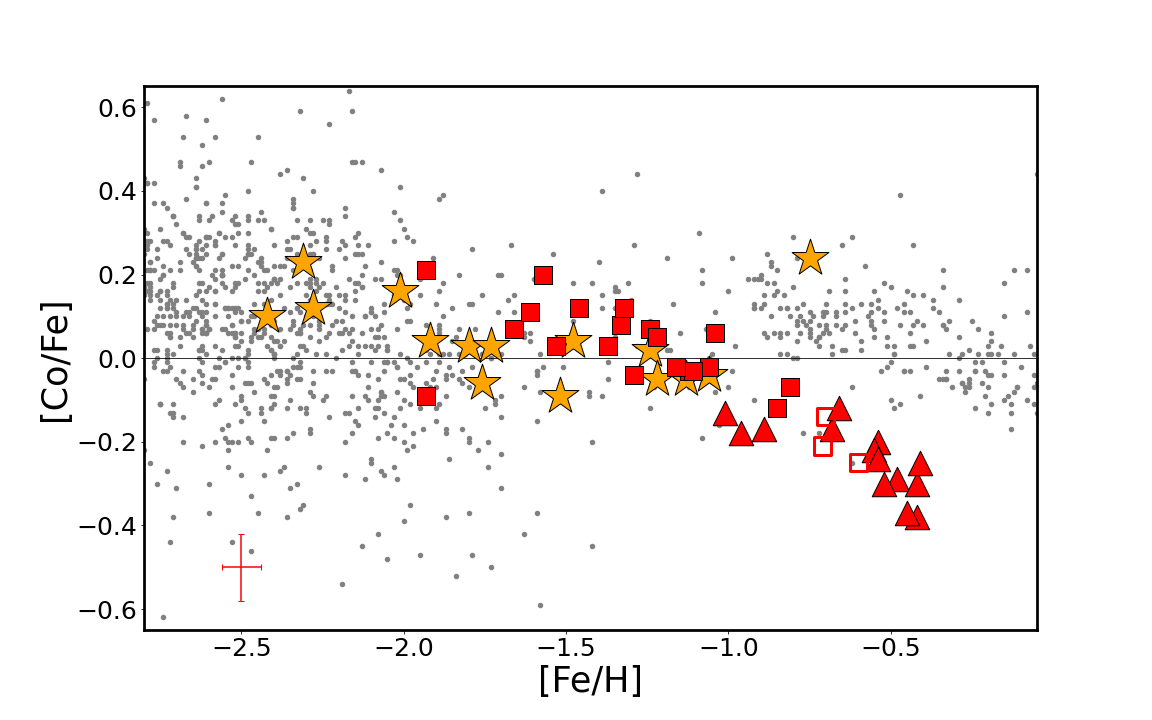}
    \includegraphics[scale=0.22]{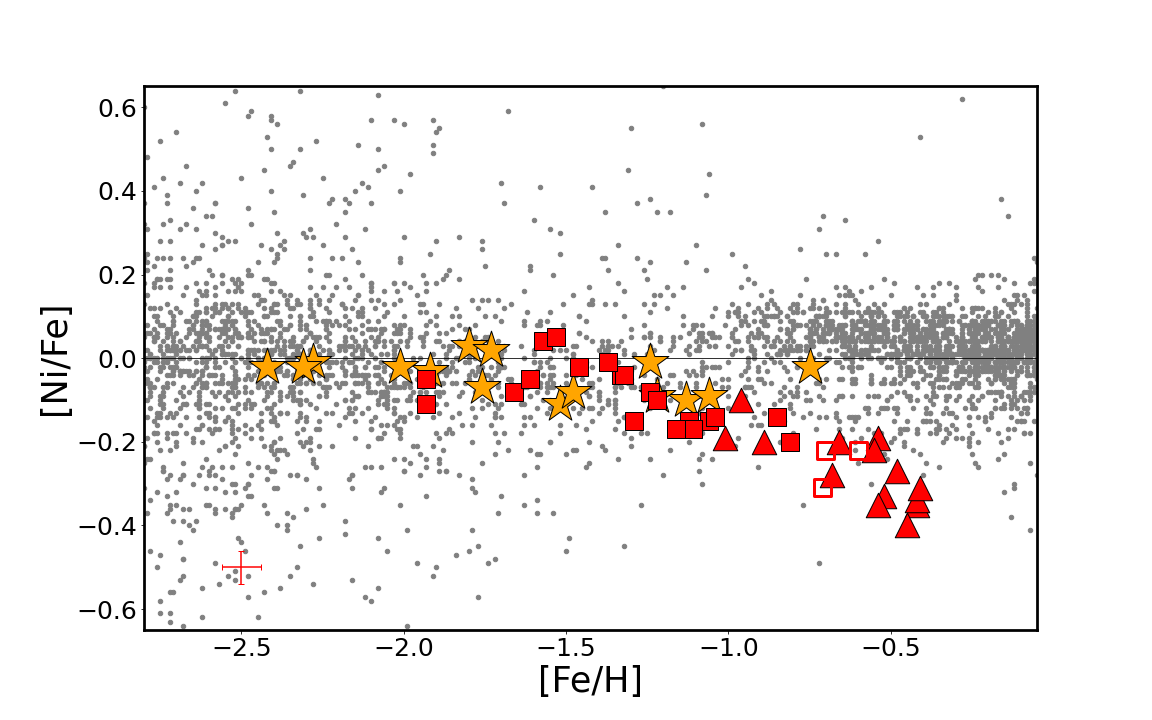}
    \includegraphics[scale=0.22]{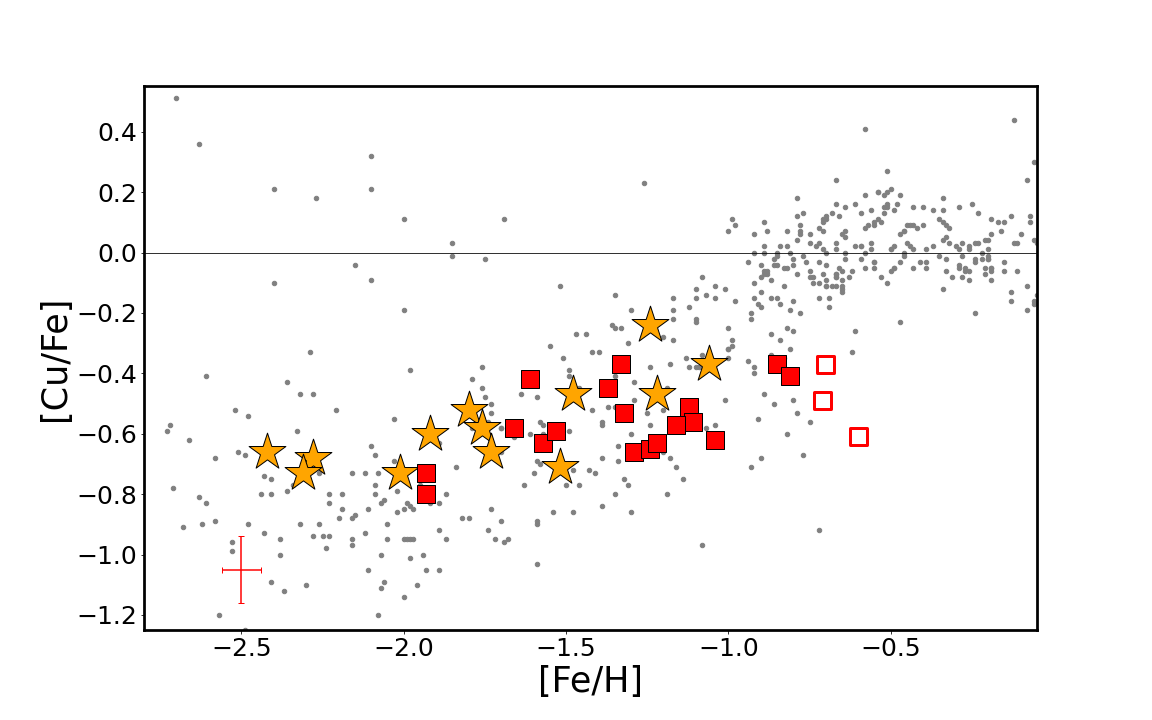}
    \includegraphics[scale=0.22]{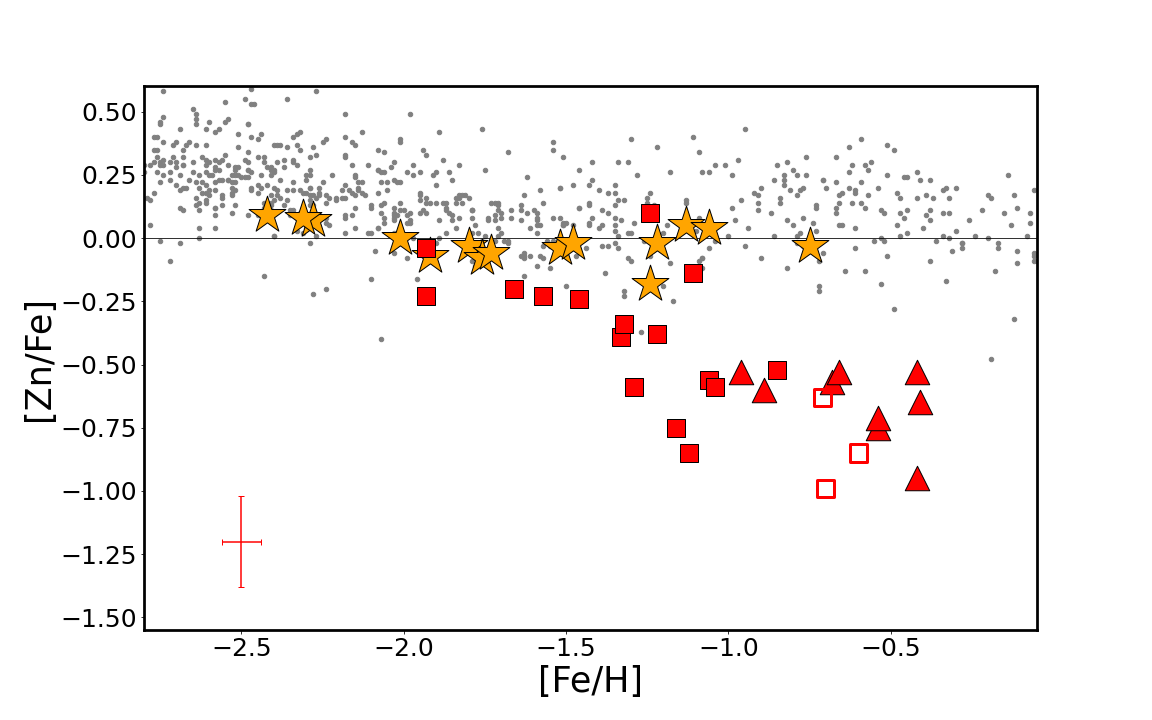}
    \caption{Behavior of [Sc/Fe], [V/Fe], [Cr/Fe] and [Mn/Fe] as a function of [Fe/H],
    same symbols of Fig~\ref{oddz}.}
    \label{ironp}
\end{figure*}

The iron-peak group includes elements formed from different nucleosynthesis paths, 
i.e. mainly massive stars (e.g. Sc, Cu, Zn), SNe Ia (e.g. Mn, Ni) or a mix of both (e.g. V, Cr, Co). 
The trends observed in Sgr dSph for this group of elements are shown in  Fig.~\ref{ironp}. 
Several of these elements (Sc, Co, Mn, Ni, Zn, Cu) show quite different chemical evolution 
pathway relative to what is observed in the Galaxy.

While for Sc and Co our theoretical knowledge is quite limited 
due to the pathological underestimation by models of observed trends across metallicities \citep{romano2010quantifying,kobayashi2020origin}, 
for Mn, Ni, Zn and Cu \citep[elements not included neither in studies based on APOGEE, 
neither in the sample by][]{vitali24arx} 
several comments can be made in the light of the different nucleosynthetic 
patterns exhibited by different progenitor sources. 

Manganese is mainly produced by SNe Ia \citep{romano2010quantifying,kobayashi2020origin,palla21}. 
The rise of [Mn/Fe] with increasing [Fe/H] reflects the small amount of Mn produced in massive stars and 
the subsequent contribution by SN~Ia (that also increases with metallicity).
In Sgr, [Mn/Fe] resembles such a trend, but with a generally higher level of abundance 
for [Fe/H]$< -$1 dex and a slower increase at higher metallicities relative to the MW.
Before the onset of SNe Ia the production of Mn is driven by CC-SNe, with the 
hypernovae\footnote{These events represent a class of massive ($\geq$ 20 $\rm M_{\odot}$) CC-SNe with 10 times or more higher explosion
energies relative to standard CC/Type II SNe ($E=10^{51}$ erg, e.g. \citealt{umeda2002,kobayashi06}). 
They are often associated with long gamma-ray burst progenitors (\citealt{nomoto13} and references therein)} 
(HNe) invoked to explain the very small [Mn/Fe] ratio at low metallicities \citep{kobayashi06,romano2010quantifying}. 
Here, the higher [Mn/Fe] level in Sgr at low metallicities, namely at [Fe/H] below that of the $\alpha$-knee (see Sect.~\ref{sec:alphas}), 
relative to the MW seems to suggest a lower contribution by HNe in Sgr.

Also Nickel is produced in very important amounts by SNe Ia
\citep{leung18,leung20,kob_ln20,palla21}. Several studies pointed out that Ni is sensitive to the white dwarf progenitors of SN~Ia
\citep[see e.g. Fig.10 and 11 in ][]{kob_ln20}, namely of near- and sub-Chandrasekhar masses (near-$M_{Ch}$ and sub-$M_{Ch}$), 
with the former producing a large amount of Ni and the latter providing sub-solar Ni abundances (see Fig. 4 in \citealt{palla21}).
Looking at the trends shown in Figure \ref{ironp}, we note that in the MW [Ni/Fe] exhibits a solar and constant value over 
the entire range of metallicities, which suggest a concurrent contribution by different near-$M_{Ch}$ and sub-$M_{Ch}$ SNe Ia progenitors \citep{palla21}.

On the other hand, Sgr stars show a decrease of [Ni/Fe] from solar (MW-like) values down to $\sim$--0.4 dex, with the drop starting 
right at the metallicity of the $\alpha$-knee, i.e. at [Fe/H]$\sim$--1.5 dex. 
A similar trend but less steep is found also by \citet{hasselquist2021apogee}, see Fig.~\ref{appendix:hass}.
This trend suggests a larger contribution of sub-Chandrasekhar mass SNe Ia in Sgr relative to the MW, providing a lower amount of Ni 
with respect to the produced Fe. 
Such a trend is also in agreement with what is seen in Figure \ref{ironp} for Mn, with a shallower rise in [Mn/Fe] observed in Sgr 
relative to the MW. In fact, Mn shares a common nucleosynthetic pattern to Ni in different SNe Ia progenitors, with near-$M_{Ch}$ 
progenitors more prone to produce Mn relative to sub-$M_{Ch}$ ones.
In general, it is worth noting that most of the classical dSphs display evidence that the dominant explosion mechanism 
of SNe Ia in these galaxies arises from sub-$M_{Ch}$ progenitors \citep{kirby19,delosreyes}.
Therefore, Sgr trends perfectly fit within this scenario.

Concerning Zinc, this element is theoretically explained by a predominant production by HNe. As these are associated with stars with very large initial stellar mass 
($\geq$ 20 $\rm M_{\odot}$), [Zn/Fe] can be a powerful tracer of the contribution 
of very massive stars and of an initial mass function (IMF) skewed against massive stars in galaxies experiencing inefficient star formation
\citep[see e.g.][]{jerabkova2018}.
Figure \ref{ironp} shows that Sgr stars exhibit a strong decline of [Zn/Fe] with increasing [Fe/H], moving from 
sub-solar values at low metallicity, already slightly lower than those measured in MW stars, down to almost [Zn/Fe]$\sim$--1 dex 
in the most metal-rich stars. 
This very steep decline can be explained as due to the  contribution of SNe Ia (that do not contribute 
to Zn production) combined with a much lower pollution by very energetic events from massive stars as HNe, which also explain 
the lower level of the plateau at low metallicities.
It is worth noting that such a trend is consistent both with the theoretical explanation proposed above for Mn and 
with the [Zn/Fe] trends observed in other MW satellites. 
In fact, [Zn/Fe] similarly drops in the most metal-rich field stars of Sculptor dSph \citep{skuladottir2017zinc}
and in the globular clusters of the SMC \citep{mucciarelli23b}.
This may indicate some level of deficiency of very massive stars in these dwarfs and, therefore, the prevalence of a top-light IMF in systems that experience a low-level star formation activity.
Finally, we highlight a large star-to-star scatter in [Zn/Fe]. This scatter is compatible with the 
high uncertainties in this abundance ratio due to the continuum location around the only available Zn transition 
at 4810 \AA\ that is located in the bluest part of the adopted UVES set-up where the S/N ratio significantly drops 
(S/N$\sim$10-20).

The last element of this group for which we note interesting differences in the observed trends 
is Copper (see Figure \ref{ironp} bottom left panel). 
To derive the Copper abundance we use the line at $\rm \lambda=5105 \AA$ that starts to be saturated 
in spectra of cool giants for [Fe/H]$>$--1 dex. Therefore, for most of the metal-rich stars of our sample 
the Cu line at 5105 \AA\ is not sensitive enough to abundance variations to derive 
reliable abundance values. For the other stars, Sgr exhibits $\rm [Cu/Fe]$ abundances significantly 
sub-solar with a mild enhancement with increasing the metallicity, 
that highly consistent resembles the behavior expected theoretically for this element, hose production is favored in presence of more metal seeds.
Cu is mainly produced from stars with $\rm M \gtrsim 8M_{\odot}$ through the so-called weak slow neutron-capture process 
\citep{romano2007_cu,prantzos2018}. 
The measured, lower [Cu/Fe] values possibly suggest a lower contribution by massive stars to the chemical enrichment 
of Sgr, in agreement with Mn and Zn trends. This behavior resembles that observed in Omega Centauri \citep{cunha02}, 
in the LMC \citep{vds13} and in the SMC \citep{mucciarelli23a}.

\subsection{Slow neutron-capture elements}

\begin{figure*}[!h]
    \centering
    \includegraphics[scale=0.22]{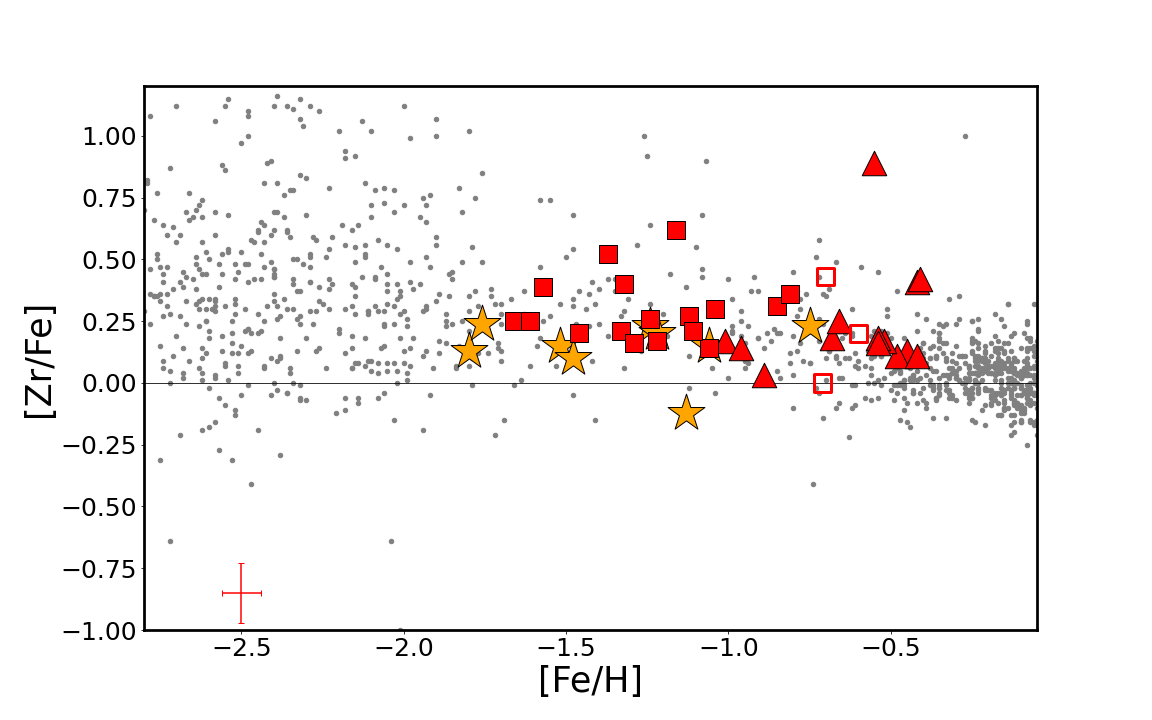}
    \includegraphics[scale=0.22]{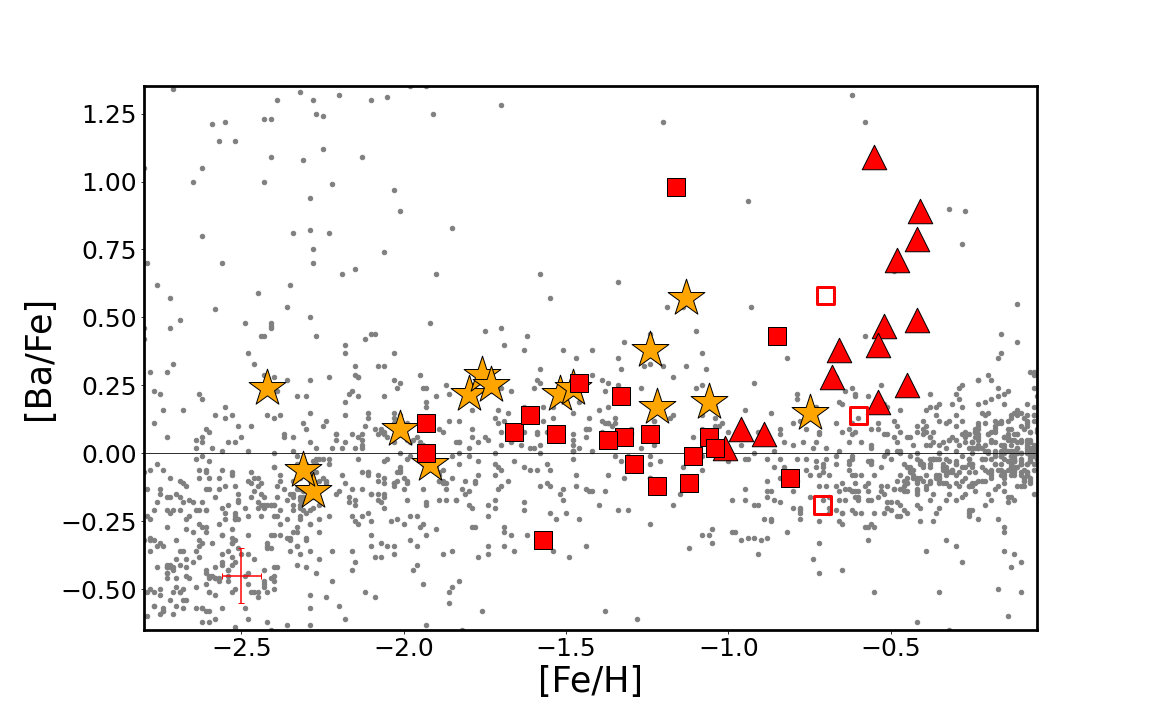}
    \includegraphics[scale=0.22]{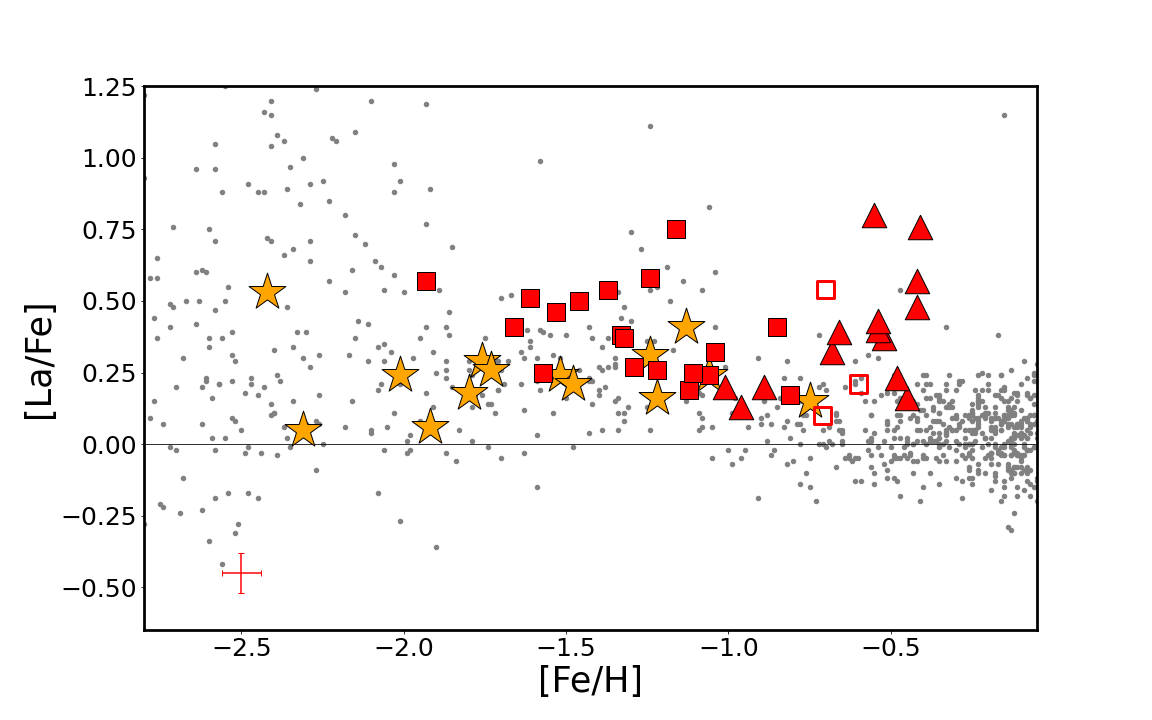}
    \includegraphics[scale=0.22]{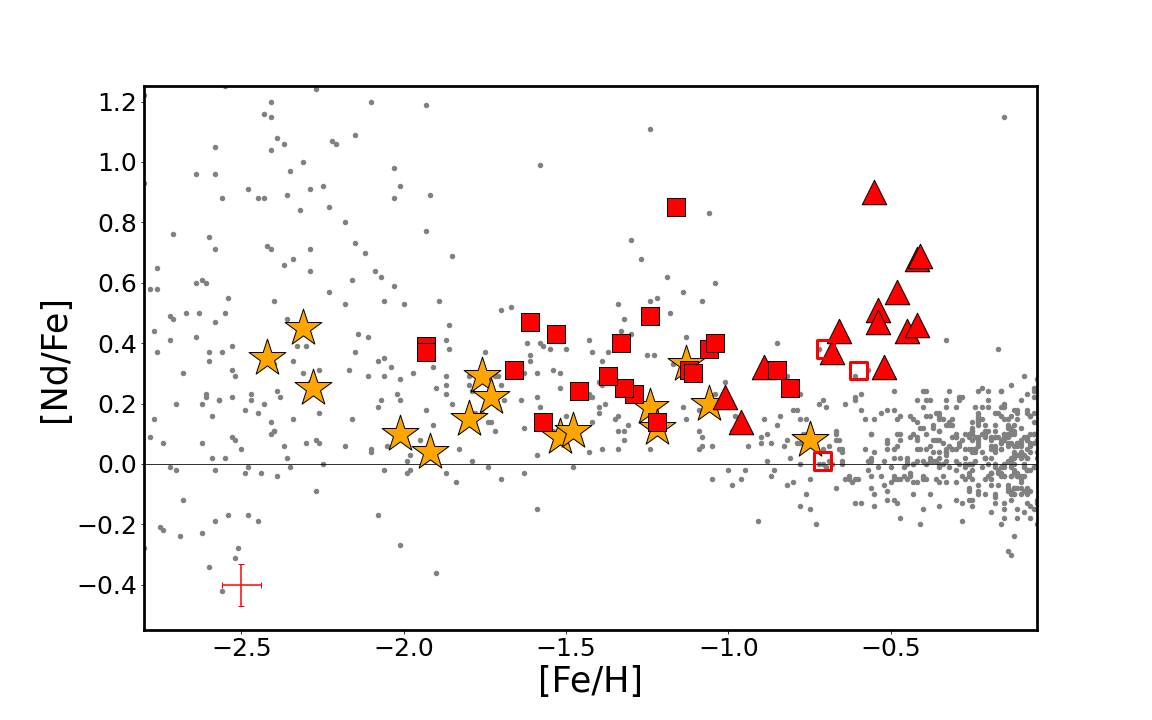}
    \includegraphics[scale=0.22]{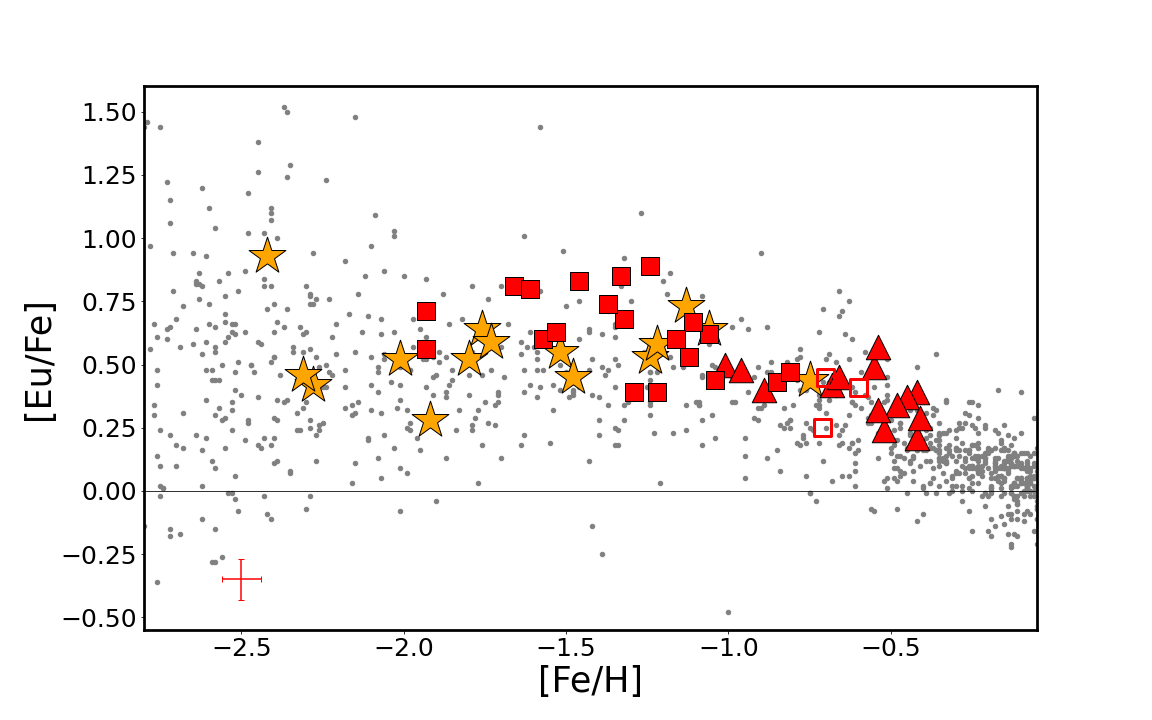}
    \caption{Behavior of [Zr/Fe], [Ba/Fe], [La/Fe] [Nd/Fe] and [Eu/Fe] as a function of [Fe/H],
    same symbols of Fig~\ref{oddz}.}
    \label{ncap}
\end{figure*}

The derived abundance ratios for Zr, Ba, La, and Nd as a function of [Fe/H] are shown in Fig. \ref{ncap}. 
These elements are mainly produced through slow neutron-capture process (hereafter, s-process) 
in stars with different ranges of masses.
In particular, Zr is significantly produced in both massive stars through the weak s-process and in AGB stars 
over a large mass range \citep[see also][]{prantzos2018}.
On the other hand, Ba, La and Nd are produced in AGB stars
with masses lower than $\sim$3--4$M_{\odot}$ \citep{gallino98,busso99,cristallo15}.
Moreover, at low metallicities significant contribution to the element budget is brought 
by the rapid neutron-capture process (hereafter, r-process, \citealt{truran81}), that occurs in rare and 
energetic events such as neutron star mergers or peculiar classes of CC-SNe (see Section~\ref{sec_r}). 
This contribution is particularly relevant for Nd, for which different studies 
\citep[e.g.][]{sneden2008,prantzos2020} attribute up to 40\% of r-process contribution.

Looking at Figure \ref{ncap}, the Sgr stars display values for these abundance ratios compatible with those 
in the MW control sample at low-intermediate metallicity, with a significant increase at [Fe/H]$> -$0.7 dex, 
In particular, Ba, La and Nd abundances reach values of about +1 dex. This value is reached following 
a clear trend at high-metallicities, which in turn should exclude that these abundances are only the effect 
of AGB mass transfer in binary systems \citep[Ba or CH stars, see][]{cseh2018,stancliffe2021}.
The large increase of [s/Fe] ratios can be attributed to a major contribution from low- to intermediate-mass stars
to the late chemical enrichment in the Sgr, which can be originated by a galaxy-wide IMF skewed towards lower masses 
relative to the canonical IMF adopted for the MW field \citep{kroupa1993}.
Though the large theoretical uncertainties still residing in the nucleosynthesis of these elements 
prevent us from drawing strong statements, we note that the above findings are consistent with the ones obtained 
from the analysis of the behavior of other key IMF tracers (most notably, Mn and Zn, see Sect.~5.3).

For Zr instead, a constant/decreasing trend is observed at high-metallicities, basically 
following the Galactic pattern. This is due to the different types of progenitors for this element. 
In fact, massive stars contribute less to the chemical enrichment of Sgr relative to the Galaxy, thus 
erasing out the greater contribution by low-intermediate mass stars.

\subsection{Rapid neutron-capture elements}
\label{sec_r}

R-process elements, like Eu measured in our sample, are thought to be produced by peculiar classes of CC-SNe, 
like collapsars (e.g. \citealt{siegel09}) and proto-magnetars/magneto-rotational SNe 
\citep[MRD-SNe, e.g.][]{winteler2012,nishimura15}, and by neutron stars mergers 
\citep[NSM,][]{lattimer74,argast2004,matteucci2014}.
The latter occur following a delay-time-distribution (similarly to SN Ia) and therefore on longer timescales 
relative to CC-SNe, and they are rare events but very efficient in producing r-process elements as suggested
by the large star-to-star scatter observed among the Milky Way metal-poor stars \citep[e.g.][]{cescutti2015}.
However, different studies agree on the need for a mixture of prompt source (i.e. CC-SNe type) and NSM 
to reproduce the Galactic [Eu/Fe] distribution  \citep{cote2019,molero2023origin}.\\

The trend of [Eu/Fe] with metallicity in Sgr stars (bottom panel of Figure \ref{ncap}) resembles that observed 
in the MW, with enhanced [Eu/Fe] values for [Fe/H]$\lesssim -$1.2/$-$1.0  dex and a subsequent decrease 
at higher metallicities, due to the onset of SNe Ia that do not produce r-process elements.
However, it is worth noting that, despite a large star-to-scatter, the metal-poor Sgr stars have [Eu/Fe] systematically 
higher by $\sim$0.25 dex than the [Eu/Fe] measured in the MW reference clusters. 
Moreover, the plateau in the [Eu/Fe] abundances is slightly more extended towards higher metallicity relative 
to the one for $\alpha$-elements. 
While our stars show enhanced [Eu/Fe] at any metallicity,  \citet{sestito24} and \citet{ou25arx} 
provide contradictory results concerning the metal-poor regime. \citet{sestito24} found solar [Eu/Fe] 
values for all their targets (5 with [Fe/H]$<$--2.2 dex and 1 with [Fe.H]$\sim$--1.2 dex), while 
the sample of \citet{ou25arx} exhibits a large star-to-star scatter, comparable to that observed 
in MW stars at similar [Fe/H] and attributed to the stochastic nature of the production sites of r-process elements. 
 Despite the different metallicity range, our results better agree with those by \citet{ou25arx}, 
in particular about the significant star-to-star scatter in [Eu/Fe]. However, it could be interesting to understand 
the origin of the discrepancy between \citet{sestito24} and \citet{ou25arx}.

The trend shown in Figure \ref{ncap} highlights that Sgr is characterized by a very efficient r-process production, 
more than what is observed in the Galaxy.
This is also observed in several galaxies of the Local Group, i.e. the SMC \citep{mucciarelli23b}, the 
LMC \citep{mucciarelli08,vds13} and classical dwarf spheroidal galaxies \citep{letarte2010,reichert20}. 
This feature of extragalactic stars has also been highlighted in the comparison of the MW halo in-situ and accreted stars 
and clusters \citep{ernandes2024,monty24}, where the latter exhibit larger [Eu/$\alpha$] values. 
Therefore, the results of this work support the idea of a strong efficiency of r-process production in MW satellites.

\section{Conclusions}
\label{conclusions}

In this work, we  describe the chemical abundances of stars in the main body of Sgr~dSph 
in the metallicity range --2.0$<$[Fe/H]$<$--0.4 dex with high resolution spectroscopy.
The chemical composition of this sample has been compared with the abundance ratios of MW GCs analyzed with the 
same assumptions of the analysis of Sgr stars, minimizing the systematics affecting the comparison between 
different chemical analyses.

In the analyzed [Fe/H] range we can identify the transition between the enrichment phase dominated by CC-SNe and 
that dominated by SNe~Ia.
In particular, the [$\alpha$/Fe] ratios suggest an $\alpha$-knee 
occurring at about [Fe/H]$\sim -$1.5/$-$1.3 dex, compatible with the lower star formation efficiency of this galaxy relative to the MW \citep{mucciarelli2017chemical} in the context of the time-delay model \citep[e.g.][]{matteucci2012}. 

At lower metallicities, Sgr stars exhibit a chemical composition compatible with MW stars of similar [Fe/H]. 
The only relevant exceptions are [Mn/Fe], [Zn/Fe] and [Eu/Fe] which display, respectively, higher (Mn, Eu) and lower (Zn) 
ratios than in MW stars. 
At higher [Fe/H], instead, the chemical patterns of Sgr significantly deviate from those of the MW for almost 
all the elements object of this study. This transition start at [Fe/H]$\sim -$1.5/$-$1.3 dex, in agreement
with the knee observed for the $\alpha$-elements.

In general, the abundance patterns of Sgr stars point to
a lower contribution by massive stars (in particular, those exploding as hypernovae, 
as revealed by [Zn/Fe] and [Mn/Fe]), a higher contribution by sub-Chandrasekhar progenitors of SNe Ia
(as revealed by [Ni/Fe] and [Mn/Fe]) and a high production efficiency of neutron-capture elements, both r-process 
([Eu/Fe]) at low metallicities and s-process ([Ba, La, Nd/Fe]) at high metallicities.
All these findings are in line with the individual elemental trends observed in other dwarf MW satellites (e.g. \citealt{kirby19,reichert20,mucciarelli23b}).
We highlight the importance of deriving abundances for some key elements that have not received the attention they deserve, 
such as, for instance, Mn, Ni and Zn.
Moreover, this study highlights the importance of pursuing the goal of a complete (in metallicity) and homogeneous 
(in derivation) sampling of the different elemental abundance patterns in these galaxies. 
In fact, it is only in this way that one can hope to break the degeneracies that affect the interpretation of the 
origin of the observed patterns, leading to a confirmation (or rejection) of our theoretical expectations.

\begin{acknowledgements}
 We thank the anonymous referee for useful comments and suggestions.
A.M., M.B., M.P. and D.R. acknowledge support from the project "LEGO – Reconstructing the building blocks of the Galaxy by chemical tagging" (P.I. A. Mucciarelli). granted by the Italian MUR through contract PRIN 2022LLP8TK\_001. 
The authors thank P. Bonifacio, E. Caffau and D. Hatzidimitriou for the help, comments and suggestions.
This project has received funding from the European Union’s Horizon 2020 research and innovation programme under the Marie Skłodowska-Curie grant agreement No 101072454, as part of the Milky Way-\textit{Gaia} Doctoral Network (\url{https://www.mwgaiadn.eu/}).
This work has made use of data from the European Space Agency (ESA) mission {\it Gaia} (\url{https://www.cosmos.esa.int/gaia}), 
processed by the {\it Gaia} Data Processing and Analysis Consortium (DPAC, \url{https://www.cosmos.esa.int/web/gaia/dpac/consortium}). 
Funding for the DPAC has been provided by national institutions, in particular the institutions participating in the {\it Gaia} Multilateral Agreement. 
Funded by the European Union (ERC-2022-AdG, "StarDance: the non-canonical evolution of stars in clusters", Grant Agreement 101093572, PI: E. Pancino). Views and opinions expressed are however those of the author(s) only and do not necessarily reflect those of the European Union or the European Research Council. Neither the European Union nor the granting authority can be held responsible for them.

\end{acknowledgements}

\bibliographystyle{aa.bst} 
\bibliography{bibliography.bib}

%
%

\appendix

\section{Tables}
\label{appendix:tables}

\begin{table*}
\centering
\caption{Sagittarius spectroscopic targets: photometric data, atmospheric parameters, [Fe/H] and RVs. 
Errors on the radial velocities are 0.1 km/s for every star. Last column lists the identification number 
of the stars in common with \citet{minelli2021homogeneous}.}
\label{tab:combined_sgr}
\scriptsize 
\begin{tabular}{ccccccccccccc}
\hline \hline    
\textit{Gaia} EDR3 ID & R.A. & Dec & $G$ & $G_{BP}$ - $G_{RP}$ &
$\rm T_{eff}$ & $\rm \log g$ & $\rm v_{turb}$ & $\rm RV$ & $\rm [Fe/H]$ & $\rm ID_{\cite{minelli2021homogeneous}}$\footnote{This columns represent the source ID used in \cite{minelli2021homogeneous} for the listed targets. \textit{We also include the source ID used internally in this work, the latter starting with "I" followed by a number. Those targets indicated in this column just as a number are the stars from \cite{minelli2021homogeneous}.}}\\ 

& (deg) & (deg) & (mag) & (mag) & ($K$) & ($cm/s^2$) & ($km/s$) & ($km/s$) & (dex) \vspace{0.1cm} \\ 

\hline 

6761207100159842048 & 283.81293381264    & -30.24403088492 & 14.8569 & 1.8329 & 4047 & 0.60 & 1.99 & 142.7 & -0.70 $\pm$ 0.06 & -- \\ 
6760463585457890688 & 283.94552443528    & -30.12538619820 & 15.2730 & 1.6697 & 4258 & 0.91 & 1.92 & 119.4 & -1.29 $\pm$ 0.07 & -- \\ 
6761215930612980992 & 283.95587588907    & -30.06320985090 & 15.5292 & 1.5574 & 4421 & 1.10 & 1.88 & 135.2 & -1.46 $\pm$ 0.06 & -- \\ 
6760462868229020544 & 283.96118989989    & -30.16519988659 & 15.5773 & 1.5426 & 4444 & 1.14 & 1.87 & 133.1 & -1.33 $\pm$ 0.06 & --  \\ 
6761218816830743936 & 283.84798568210    & -30.00368432035 & 15.5768 & 1.4647 & 4569 & 1.20 & 1.85 & 152.9 & -1.93 $\pm$ 0.08 & --  \\ 
6760462829543616640 & 283.96000062282    & -30.17971311810 & 15.6083 & 1.5294 & 4465 & 1.16 & 1.86 & 127.2 & -0.71 $\pm$ 0.06 & --  \\ 
6760402463806832000 & 283.73051807680    & -30.81065919159 & 14.6014 & 1.9539 & 3909 & 0.40 & 2.04 & 155.4 & -1.57 $\pm$ 0.06 & -- \\ 
6760305328822935296 & 283.44999801454    & -30.99108124685 & 14.8026 & 1.7939 & 4095 & 0.61 & 1.99 & 158.9 & -1.32 $\pm$ 0.08 & -- \\ 
6760397687804020736 & 283.70967770804    & -30.88972547528 & 15.6506 & 1.5381 & 4451 & 1.17 & 1.86 & 138.4 & -0.60 $\pm$ 0.05 & --  \\ 
6760397546039591808 & 283.68963531518    & -30.92637738773 & 15.7147 & 1.5777 & 4390 & 1.16 & 1.86 & 139.1 & -1.24 $\pm$ 0.07 & --  \\ 
6760397619083550336 & 283.67572616798    & -30.90141349639 & 14.9833 & 1.8591 & 4016 & 0.63 & 1.99 & 133.6 & -0.85 $\pm$ 0.05 & --  \\ 
6760319931712049152 & 283.44116545413    & -30.87046267833 & 15.4209 & 1.7681 & 4128 & 0.88 & 1.93 & 158.2 & -0.81 $\pm$ 0.06 & --  \\ 
6760379648939656064 & 284.19411719184    & -30.75505902806 & 14.9094 & 1.7096 & 4204 & 0.73 & 1.96 & 150.3 & -1.66 $\pm$ 0.08 & -- \\ 
6760431566506754560 & 284.12699453751    & -30.60599375853 & 15.0498 & 1.6462 & 4291 & 0.84 & 1.94 & 133.0 & -1.93 $\pm$ 0.09 & -- \\ 
6760382225920034304 & 284.29779457445    & -30.65084574665 & 15.2845 & 1.6630 & 4267 & 0.96 & 1.91 & 127.3 & -1.22 $\pm$ 0.06 & -- \\ 
6760381435646033152 & 284.27803818895    & -30.70157648060 & 14.9226 & 1.8520 & 4025 & 0.61 & 1.99 & 129.7 & -1.12 $\pm$ 0.06 & --  \\ 
6760385146497860352 & 284.23783495489    & -30.61020436377 & 15.0060 & 1.9699 & 3891 & 0.55 & 2.00 & 142.8 & -1.06 $\pm$ 0.06 & --  \\ 
6760431429057147264 & 284.09167379491    & -30.62466021796 & 15.0966 & 1.9477 & 3915 & 0.60 & 1.99 & 155.3 & -1.11 $\pm$ 0.05 & --  \\ 
6761202908271029632 & 283.37749031930    & -30.12314423999 & 14.8856 & 1.6773 & 4247 & 0.74 & 1.96 & 132.6 & -1.61 $\pm$ 0.08 & -- \\ 
6761196684833413760 & 283.40172879410    & -30.28196402940 & 15.6179 & 1.5786 & 4389 & 1.12 & 1.87 & 141.9 & -1.37 $\pm$ 0.06 & --  \\ 
6761177825661827584 & 283.32421439481    & -30.40973596635 & 15.6909 & 1.5132 & 4490 & 1.21 & 1.85 & 127.1 & -1.53 $\pm$ 0.07 & --  \\ 
6761179646727792256 & 283.48541432771    & -30.45572417837 & 14.8143 & 2.0222 & 3836 & 0.43 & 2.03 & 158.1 & -1.16 $\pm$ 0.06 & --  \\ 
6761178199294010240 & 283.33504724499    & -30.39179625344 & 15.1433 & 1.9329 & 3932 & 0.63 & 1.99 & 129.5 & -1.04 $\pm$ 0.05 & --  \\ 
6760412015814104576 & 283.94469583595    & -30.59023509442 & 15.0906 & 1.8690 & 4023 & 0.68 & 1.97 & 147.1 & -1.01 $\pm $0.05 & 2300127 \\
6760448712016440320 & 283.87831943063    & -30.47218992929 & 15.6683 & 1.8832 & 4014 & 0.91 & 1.92 & 148.1 & -0.42 $\pm $0.06 & 2300196 \\
6760425141234915328 & 283.82984880442    & -30.50784695524 & 15.8746 & 1.8697 & 4037 & 1.00 & 1.90 & 154.9 & -0.45 $\pm $0.06 & 2300215 \\
6760423904283773440 & 283.73281303667    & -30.54539351576 & 15.6178 & 1.9604 & 3938 & 0.83 & 1.94 & 131.6 & -0.48 $\pm $0.06 & 2409744 \\
6761177001027997184 & 283.44099151341    & -30.43046644552 & 15.8522 & 1.7598 & 4178 & 1.09 & 1.88 & 153.8 & -0.52 $\pm $0.06 & 3600230 \\
6761178203618946304 & 283.34314992525    & -30.39651246472 & 15.9289 & 1.7806 & 4110 & 1.08 & 1.88 & 156.8 & -0.54 $\pm $0.05 & 3600262 \\
6761173148412314112 & 283.43845326993    & -30.51554263531 & 15.9426 & 1.7844 & 4120 & 1.09 & 1.88 & 143.8 & -0.42 $\pm $0.06 & 3600320 \\
6760428779046508928 & 283.74285068159    & -30.47235023209 & 15.5165 & 1.8841 & 4135 & 1.15 & 1.86 & 151.8 & -0.55 $\pm $0.07 & 3800318 \\
6760429230043653888 & 283.74135535208    & -30.44872763385 & 15.7856 & 1.8413 & 4318 & 1.15 & 1.87 & 134.4 & -0.96 $\pm $0.07 & 3800558 \\
6760428508489236224 & 283.63782444343    & -30.45531956125 & 15.3743 & 1.7515 & 4188 & 0.90 & 1.92 & 156.2 & -0.68 $\pm $0.06 & 4214652 \\
6760421116820695936 & 283.50891629933    & -30.60608028041 & 15.5338 & 1.7633 & 4235 & 1.00 & 1.94 & 143.1 & -0.89 $\pm $0.06 & 4302733 \\
6761170472678163840 & 283.41927945235    & -30.5953157088  & 15.5345 & 1.7873 & 4235 & 1.00 & 1.94 & 119.9 & -0.66 $\pm $0.06 & 4304445 \\
6761170811950036864 & 283.33243024962    & -30.62788059457 & 15.8542 & 1.8207 & 4186 & 1.10 & 1.88 & 159.2 & -0.54 $\pm $0.05 & 4402285 \\
6761174698926036224 & 283.30377181571    & -30.53438431469 & 16.0643 & 1.7706 & 4170 & 1.17 & 1.86 & 144.1 & -0.41 $\pm $0.09 & 4408968 \\ 
\hline
\end{tabular}
\end{table*}

\section{Comparison with \citet{hasselquist2021apogee}}
\label{appendix:hass}

The [Mg/Fe], [Al/Fe], [Si/Fe], [Ca/Fe] and [Ni/Fe] abundance ratios measured 
in this study were compared with those obtained by the H-band spectra of the APOGEE survey 
discussed by \citet{hasselquist2021apogee}. Because the \citet{hasselquist2021apogee} sample 
includes also Sgr stars located in the streams, for this comparison we consider only 
Sgr stars outside the tidal radius of M54 (in order to avoid the contamination from cluster stars) 
and within 60' from Sgr center, similar to the spatial region where our targets are located.

\begin{figure*}
    \centering
    \includegraphics[scale=0.22]{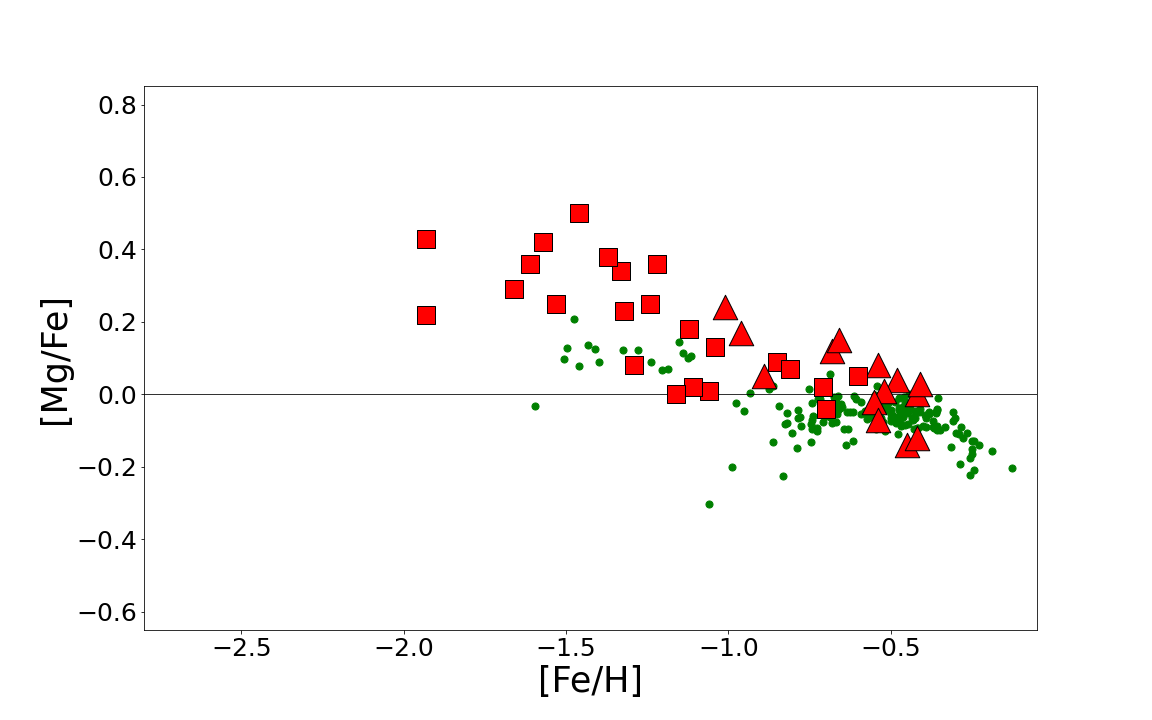}
    \includegraphics[scale=0.22]{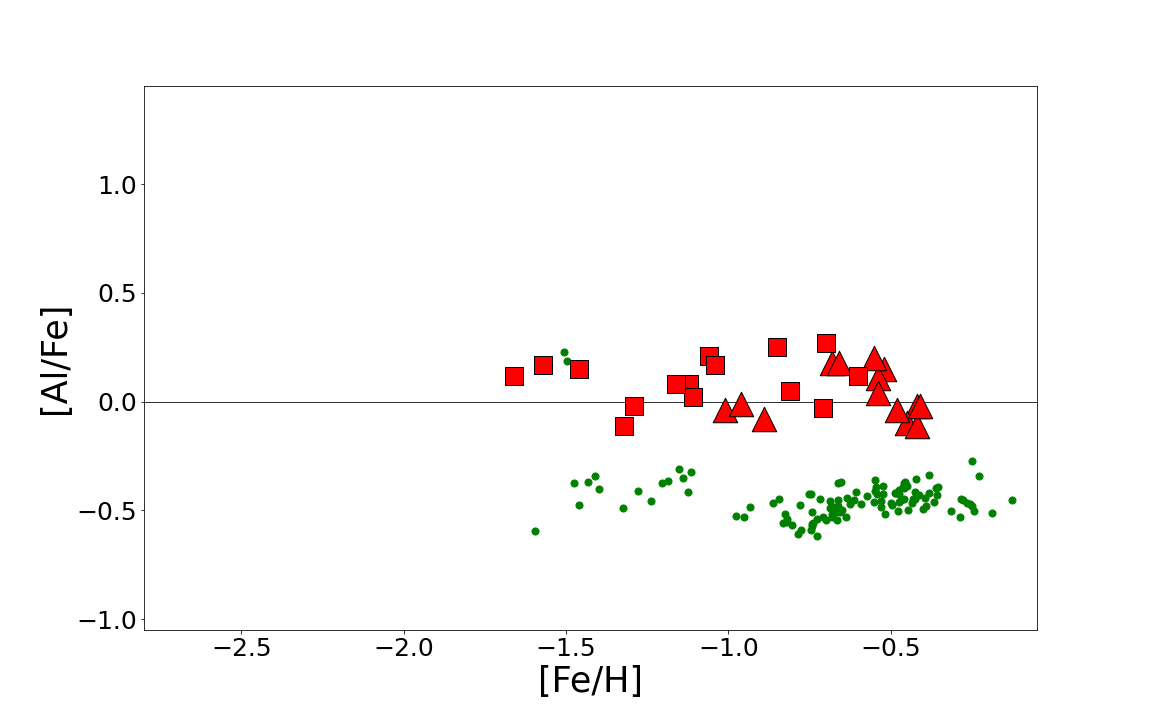}
    \includegraphics[scale=0.22]{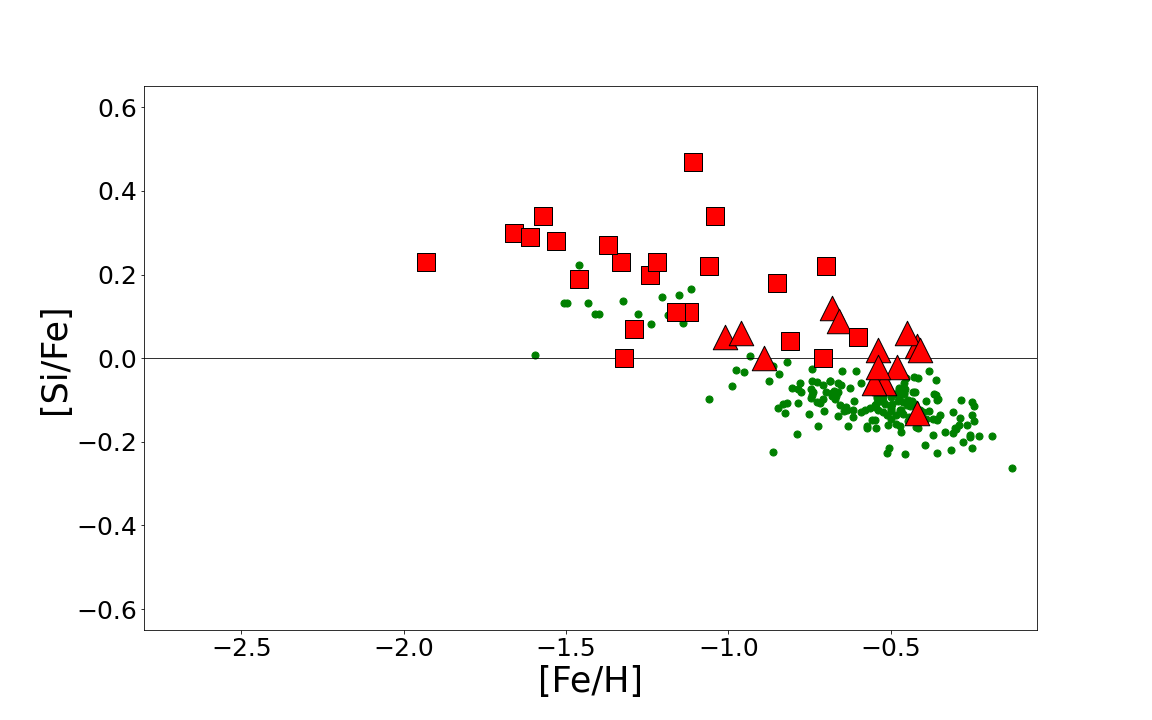}
    \includegraphics[scale=0.22]{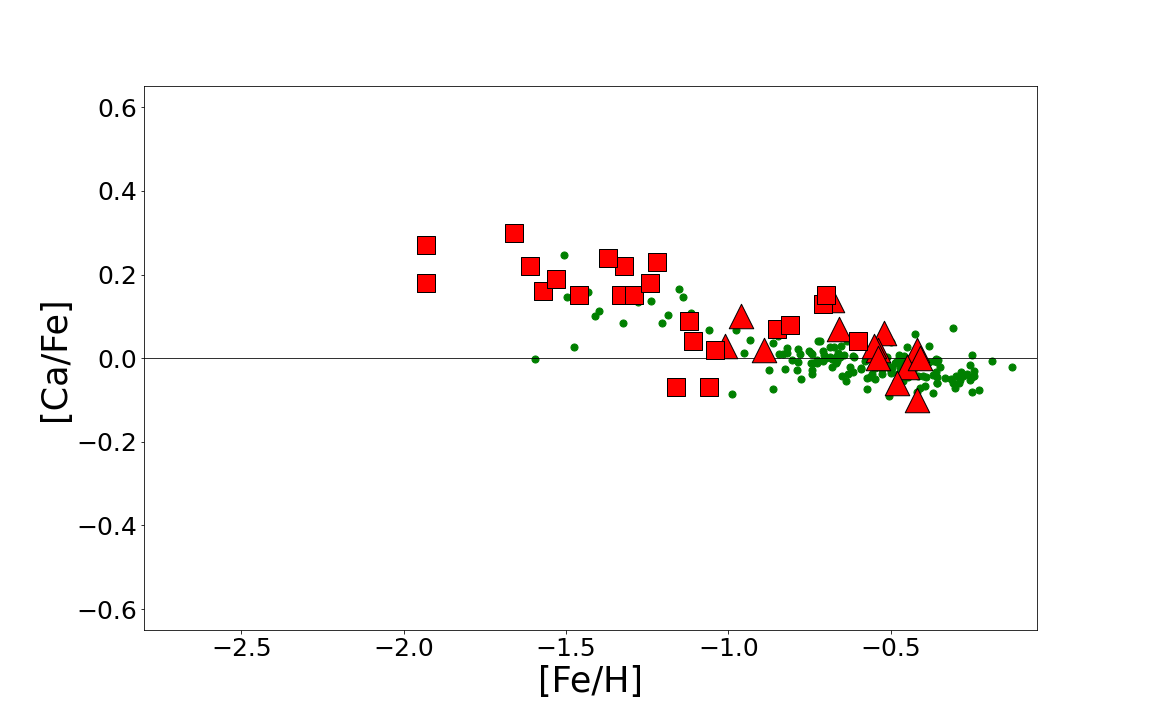}
    \includegraphics[scale=0.22]{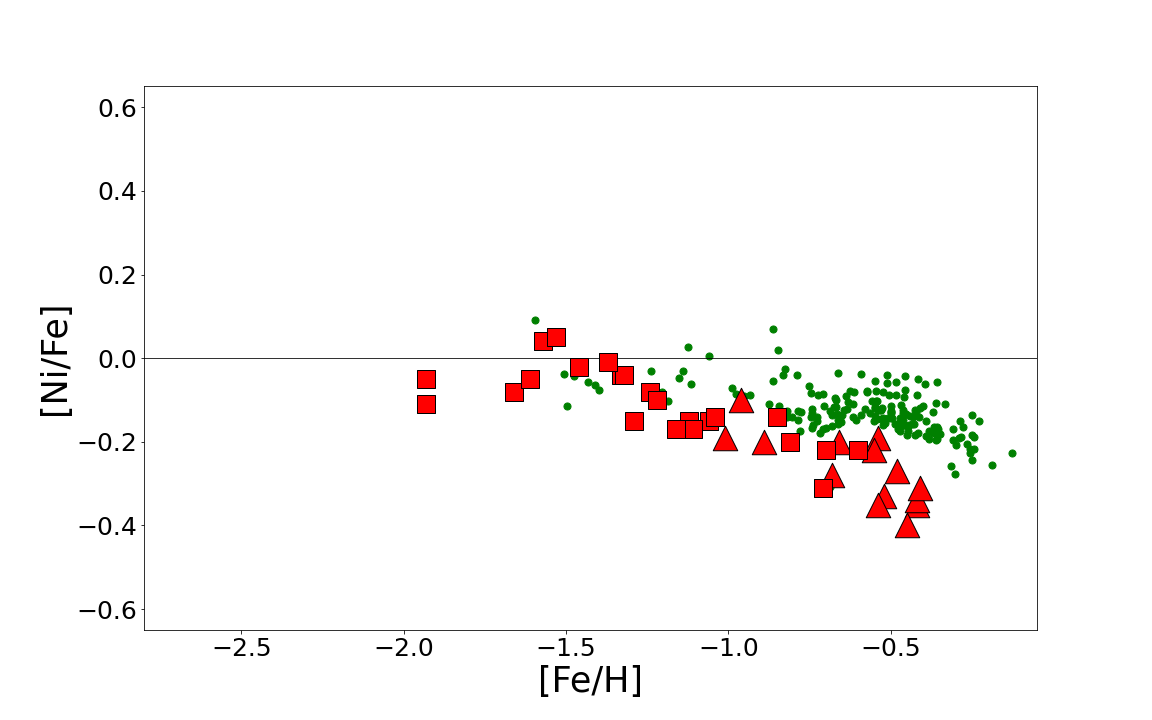}
    \caption{Comparison between this study (red symbols) and \citet[][green circles]{hasselquist2021apogee}.}
    \label{alfa-apo}
\end{figure*}

\end{document}